\documentclass[aps,prb,twocolumn,superscriptaddress,10pt]{revtex4-2}
\usepackage[utf8]{inputenc}
\usepackage{graphicx, color}
\usepackage{xcolor}
\usepackage{amsmath}
\usepackage{mathtools}
\usepackage{amssymb}
\usepackage{bm}
\usepackage{siunitx}
\usepackage{gensymb}
\usepackage{textcomp}
\usepackage[section]{placeins}
\usepackage{float}
\usepackage{longtable}
\usepackage{booktabs}
\usepackage{url}
\usepackage{nameref}
\usepackage{hyperref}
\usepackage{xspace}
\usepackage[toc,page]{appendix}
\definecolor{myblue}{RGB}{0,0,180}
\definecolor{link}{rgb}{0.1,0.1,0.9}
\hypersetup{colorlinks=true,linkcolor=link,citecolor=link,urlcolor=link,linktocpage}
\newcommand*\pct{\protect\scalebox{0.9}{\%}\xspace}
\makeatletter
\g@addto@macro\bfseries{\boldmath}
\makeatother
\DeclareRobustCommand{\rchi}{{\mathpalette\irchi\relax}}
\newcommand{\irchi}[2]{\raisebox{\depth}{$#1\chi$}}
\DeclareUnicodeCharacter{2212}{\textendash}
\newcommand{\sg}{$P$6$_{3}$/$mmc$\xspace}
\newcommand{\BSCO}{Ba$_3$Sb$_{1+x}$Co$_{2-x}$O$_{9-\delta}$\xspace}
\begin{document}
	
	\preprint{APS/123-QED}
	
	\title{Cluster Spin Glass State in Ba$_3$Sb$_{1+x}$Co$_{2-x}$O$_{9-\delta}$: Cation Disorder and Mixed-Valence Co Dimers}
	
	\author{Anzar Ali}
	\email{a.ali@fkf.mpg.de}
	\affiliation{Max Planck Institute for Solid State Research, Heisenbergstraße 1, D-70569 Stuttgart, Germany}
	
	\author{Guratinder Kaur}
	\affiliation{Max Planck Institute for Solid State Research, Heisenbergstraße 1, D-70569 Stuttgart, Germany}
	\affiliation{School of Physics and Astronomy, The University of Edinburgh, Edinburgh EH9 3JZ, United Kingdom}
	
	\author{Lukas Keller}
	\affiliation{Laboratory for Neutron Scattering and Imaging, Paul Scherrer Institute, CH-5232 Villigen, Switzerland}
	
	\author{Masahiko Isobe}
	\email{m.isobe@fkf.mpg.de}
	\affiliation{Max Planck Institute for Solid State Research, Heisenbergstraße 1, D-70569 Stuttgart, Germany}
	
	\begin{abstract}
		
		We investigate the structural, magnetic, and thermodynamic properties of \BSCO\ ($x$ = 0.04, $\delta$ = 0.54), a hexagonal perovskite featuring face-sharing CoO$_6$ octahedra that forms Co dimers. DC and AC magnetization measurements reveal a frequency-dependent spin-freezing transition consistent with glassy dynamics. AC susceptibility fits best to the Vogel-Fulcher model, indicating collective freezing of interacting spin clusters. Isothermal magnetization follows the Langevin function, suggesting finite-sized magnetic clusters rather than isolated paramagnetic moments. Non-equilibrium dynamics, evidenced by thermoremanent magnetization and memory effects, further support a spin-glass-like state. Heat capacity shows no sharp anomalies, and neutron powder diffraction confirms the absence of magnetic Bragg peaks down to 1.5~K, ruling out long-range magnetic order. Rietveld refinement reveals significant Co/Sb intersite disorder ($\sim$~30\pct) and oxygen non-stoichiometry, introducing exchange randomness and frustration that drive the spin-glass-like behavior. Electrical resistivity exhibits Arrhenius-type temperature dependence with an activation energy of 0.173~eV, consistent with semiconducting behavior. Temperature-dependent X-ray diffraction shows no structural phase transitions, confirming that the spin-glass-like state is not lattice-driven. Our results establish \BSCO\ as a cluster spin-glass candidate, where Co dimers, disorder, and geometric frustration prevent long-range order, leading to slow spin dynamics. These findings highlight the role of cation disorder and oxygen vacancies in stabilizing unconventional magnetic states in cobalt-based hexagonal perovskites.
		
	\end{abstract}  
	
	\maketitle
	
	\section{Introduction}
	\label{Intro}
	
	The study of geometrically frustrated and disordered magnetic systems remains a focal point in condensed matter physics, primarily due to their capacity to host unconventional magnetic ground states \cite{Balents2010, Broholm2020, Moessne2006}. Among these, spin glasses; materials characterized by frozen, disordered spin configurations, exhibit rich physics driven by competing exchange interactions and disorder-induced frustration \cite{Binder1986, Binder2005}. In transition-metal oxides, the interplay between local disorder, mixed-valence states, and geometric frustration give rise to complex magnetic behaviors, covering spin-glass states to reentrant magnetic phases \cite{Katukuri2015, Sahoo2018, Bridges2006}.
	
	Hexagonal perovskites of the form A$_4$BB'$_3$O$_{12}$ and A$_3$BB'$_2$O$_9$ have emerged as versatile platforms for exploring such exotic magnetism \cite{Nguyen2021}. These materials accommodate diverse structural motifs, including face-sharing B'O$_6$ octahedral trimers or dimers, fostering strong intra-unit magnetic coupling that competes with weaker inter-unit interactions. This competition, further modulated by cation disorder and charge fluctuations, can profoundly influence the magnetic ground state. For example, trimer-based compounds such as Ba$_4$NbIr$_3$O$_{12}$ \cite{Nguyen2019}, Ba$_4$NbRh$_3$O$_{12}$ \cite{Bandyopadhyay2024}, and Ba$_4$TaMn$_3$O$_{12}$ \cite{Ali2024}, along with dimer-based systems like Ba$_3$NiSb$_2$O$_9$ \cite{Quilliam2016}, Ba$_3$CuSb$_2$O$_9$ \cite{Zhou2011}, and Ba$_3$FeRu$_2$O$_9$ \cite{Middey2011}, display varying degrees of magnetic frustration and disorder, leading to behaviors that span from spin-glass states to potential quantum-disordered phases.
	
	Disorder-induced quantum spin-liquid-like states have been proposed in several frustrated magnetic systems where magnetic and nonmagnetic ions either substitute each other or share the same crystallographic site. Examples include Ho$_2$Ti$_2$O$_7$~\cite{Savary2017}, Pr$_2$Zr$_2$O$_7$~\cite{Wen2017}, Ba$_3$CuSb$_2$O$_9$~\cite{Smerald2015}, and Y$_2$CuTiO$_6$~\cite{Kundu2020}. In such cases, disorder can introduce long-range quantum entanglement, stabilizing diverse ground states such as Coulombic spin liquids, Mott glass phases, or conventional glassy states, depending on its extent~\cite{Savary2017}. In Y$_2$CuTiO$_6$, magnetic Cu$^{2+}$ ($S = 1/2$) and nonmagnetic Ti$^{4+}$ randomly share the same crystallographic site in a 50:50 ratio, forming a triangular lattice. Despite significant site dilution, no long-range order or spin freezing is observed down to 50~mK, indicating a disorder-driven cooperative paramagnetic state. Similarly, in Ba$_3$CuSb$_2$O$_9$, 50\% site sharing of Cu$^{2+}$ with Sb$^{5+}$ leads to spin-orbital liquid-like behavior~\cite{Smerald2015}. Sr$_3$CuSb$_2$O$_9$, with 33\% Cu and 67\% Sb site disorder, also shows spin-liquid-like behavior down to 65~mK~\cite{Kundu2020_2}.
	
	Controlled chemical substitution has proven an effective strategy to tune quantum states in frustrated magnets. For instance, substituting Ru$^{3+}$ with nonmagnetic Ir$^{3+}$ in the Kitaev magnet $\alpha$-Ru$_{1-x}$Ir$_x$Cl$_3$ introduces quenched disorder that stabilizes a spin-liquid-like state~\cite{Baek2020}. Such substitutions typically introduce both structural disorder—through random occupancy of crystallographic sites—and magnetic disorder—through disruption of magnetic exchange pathways. Intersite cation disorder is also known to strongly influence magnetic ground states, often promoting spin glass behavior in frustrated magnets. In MnSb$_2$Se$_4$, $\sim$~26\% antisite disorder between Mn and Sb sites leads to long-range antiferromagnetic order ($T_N = 22.5$~K) with type-II multiferroicity, whereas increasing the disorder to $\sim$~40\% transforms the system into a cluster spin glass with significantly modified magnetic and transport properties~\cite{Rahul2022}. A similar scenario occurs in the hexagonal perovskite Ba$_3$Co[Co$_{0.25}$Ru$_{0.75}$]$_2$O$_9$, where mixed-valence Co and Ru ions occupy the face-sharing octahedral sites, and the resulting cation disorder and charge transfer effects stabilize a spin glass state with a giant exchange bias~\cite{Pal2024}.
	
	Ba$_3$Sb$_{1+x}$Co$_{2-x}$O$_{9-\delta}$, the focus of this study, belongs to the family of 6H-hexagonal perovskites and features CoO$_6$ octahedral dimers arranged in a triangular network. Unlike well-ordered dimer systems that often stabilize quantum spin-singlet states or develop long-range magnetic order, \BSCO\ exhibits a cluster spin-glass ground state—an outcome of pronounced magnetic frustration and substantial cation disorder. This disorder encompasses both structural contributions, arising from Co/Sb intersite mixing ($\sim$30\%), and magnetic disruptions, due to broken dimer connectivity and randomized exchange interactions. These perturbations destabilize the uniform magnetic exchange network, giving rise to spatially inhomogeneous spin environments and resulting in a frozen, disordered magnetic state.
	
	In this context, dimer-based hexagonal perovskites with the general formula A$_3$MM$'$$_{2}$O$_{9}$ provide an ideal platform to investigate disorder- and frustration-driven glassy magnetism, as they are inherently susceptible to M/M$'$ intersite mixing. In \BSCO, the face-sharing CoO$_6$ dimers promote strong intra-dimer antiferromagnetic or ferromagnetic correlations via short Co--Co distances. However, the presence of intersite disorder and potential mixed valency fragments the magnetic lattice, suppressing coherent long-range correlations. These conditions favor the formation of weakly coupled spin clusters embedded in a disordered matrix—an ideal setting for the emergence of a cluster spin-glass state. Thus, systems like \BSCO, by virtue of their structural topology and chemical instability, are naturally predisposed to hosting unconventional frozen magnetic states, offering a fertile ground for exploring novel manifestations of frustration-induced glassy behavior.

	Experimental data from DC and AC magnetic susceptibility measurements show a bifurcation between field-cooled and zero-field-cooled magnetization below $\sim$~50~K, alongside frequency-dependent peaks in AC susceptibility, which is a hallmark of spin glass behavior. Neutron diffraction confirms the absence of magnetic Bragg peaks down to 1.5~K, ruling out the possibility of conventional long-range order. Specific heat measurements reveal no sharp anomalies, consistent with a frozen, disordered magnetic state. Scaling analysis of the AC susceptibility supports a cluster spin glass scenario, where weakly interacting ferromagnetic clusters remain dynamically frustrated.
	
	Our findings highlight that the spin glass state in \BSCO results from the interplay of cation disorder, mixed-valence states, and competing magnetic interactions. Disruptions in the ideal face-sharing dimer network inhibit singlet formation and long-range order, stabilizing a glassy state composed of frozen spin clusters. This study deepens our understanding of disorder-driven magnetism in low-dimensional oxides and underscores the complex relationship between structural disorder and magnetic behavior in transition-metal perovskites.

	\section{Experimental Details}
	\label{Exp}
	
	\subsection{Sample Preparation}
	
	Polycrystalline samples of \BSCO\ were synthesized using the conventional solid-state reaction method. High-purity BaCO$_3$ (99.997\pct, Thermo Scientific), Sb$_2$O$_5$ (99.998\pct, Thermo Scientific), and Co$_3$O$_4$ (99.9985\pct, Thermo Scientific) were used as starting materials. The precursors were weighed in stoichiometric ratios, thoroughly mixed in an agate mortar, and subjected to initial calcination at 900~$^\circ$C for 12 hours. The resulting powders were then pressed into pellets and sintered in air at 1000~$^\circ$C for 24 hours, followed by a subsequent annealing step at 1100~$^\circ$C for another 24 hours, with multiple intermediate grindings and pelletization to ensure homogeneity.
	
	\subsection{X-ray and Neutron Diffraction Measurements}
	Room-temperature X-ray diffraction (XRD) data for \BSCO\ were collected on a Rigaku SmartLab diffractometer using monochromatic Cu-K$\alpha$ radiation ($\lambda = 1.5406$~\text{\AA}), with measurements spanning $10^\circ \leq 2\theta \leq 120^\circ$ at a step size of $0.02^\circ$. Temperature-dependent XRD studies were conducted at the Centre for Science at Extreme Conditions (CSEC, University of Edinburgh, UK) using an Oxford cryostat for the measurements ranging from 300 to 15~K.
	
	Neutron powder diffraction (NPD) data were acquired at the DMC cold neutron diffractometer at SINQ, Paul Scherrer Institute (PSI, Switzerland). Measurements covered temperatures of 300, 80, 50, 30, 20, and 1.5~K, with an incident neutron wavelength of 2.454~\text{\AA}\ and angular range $10^\circ \leq 2\theta \leq 120^\circ$ (step size $0.1^\circ$). The sample was loaded in a vanadium can and cooled using a helium cryostat, achieving a base temperature of 1.5~K. 
	
	Rietveld refinements of XRD and NPD data were performed using \textit{Jana2006} \cite{Jana2006} to model crystal structures, site occupancies, and thermal parameters. Diffraction patterns were corrected for instrumental contributions and background before refinement.
	
	\subsection{Energy-Dispersive X-ray Spectroscopy}
	\label{EDS}
	
	Energy-dispersive X-ray (EDX) spectra were acquired using a NORAN System 7 (NSS212E) detector integrated into a Tescan Vega (TS-5130MM) scanning electron microscope. Semiquantitative EDX analysis of the \BSCO\ pellet yielded elemental ratios of Ba:Sb:Co = 3.00:1.05:2.10, confirming the expected stoichiometry within experimental uncertainty.
	
	\subsection{Magnetic, Thermal, and Electrical Transport Measurements}  
	
	The temperature dependence of DC magnetic susceptibility was measured under zero-field-cooled (ZFC) and field-cooled (FC) conditions using a SQUID magnetometer (Quantum Design, MPMS-XL) in applied fields ranging from 50~Oe to 10,000~Oe over the temperature range of 1.8–300~K. Isothermal magnetization measurements were performed in magnetic fields sweeping from $-70$~kOe to $+70$~kOe. AC susceptibility was measured between 2 and 50~K with an AC magnetic field of 10~Oe amplitude and frequencies varying from 1 to 1000~Hz using the same SQUID magnetometer.
	
	Specific heat capacity was measured using the thermal relaxation method in a Quantum Design PPMS over the temperature range of 1.8–200~K. The sintered pellet was affixed to a thin alumina platform with Apiezon-N grease to ensure optimal thermal contact.
	
	Electrical resistivity was measured on sintered rectangular pellets using a four-probe method in a Quantum Design PPMS over the temperature range of 120–300~K. Gold wires and silver epoxy were used for electrical contacts. The measured resistance was converted to resistivity based on the sample’s dimensions.
	
	\section{Results}
	\label{Result}
	
	\begin{figure*}
		\includegraphics[width=1.0\linewidth]{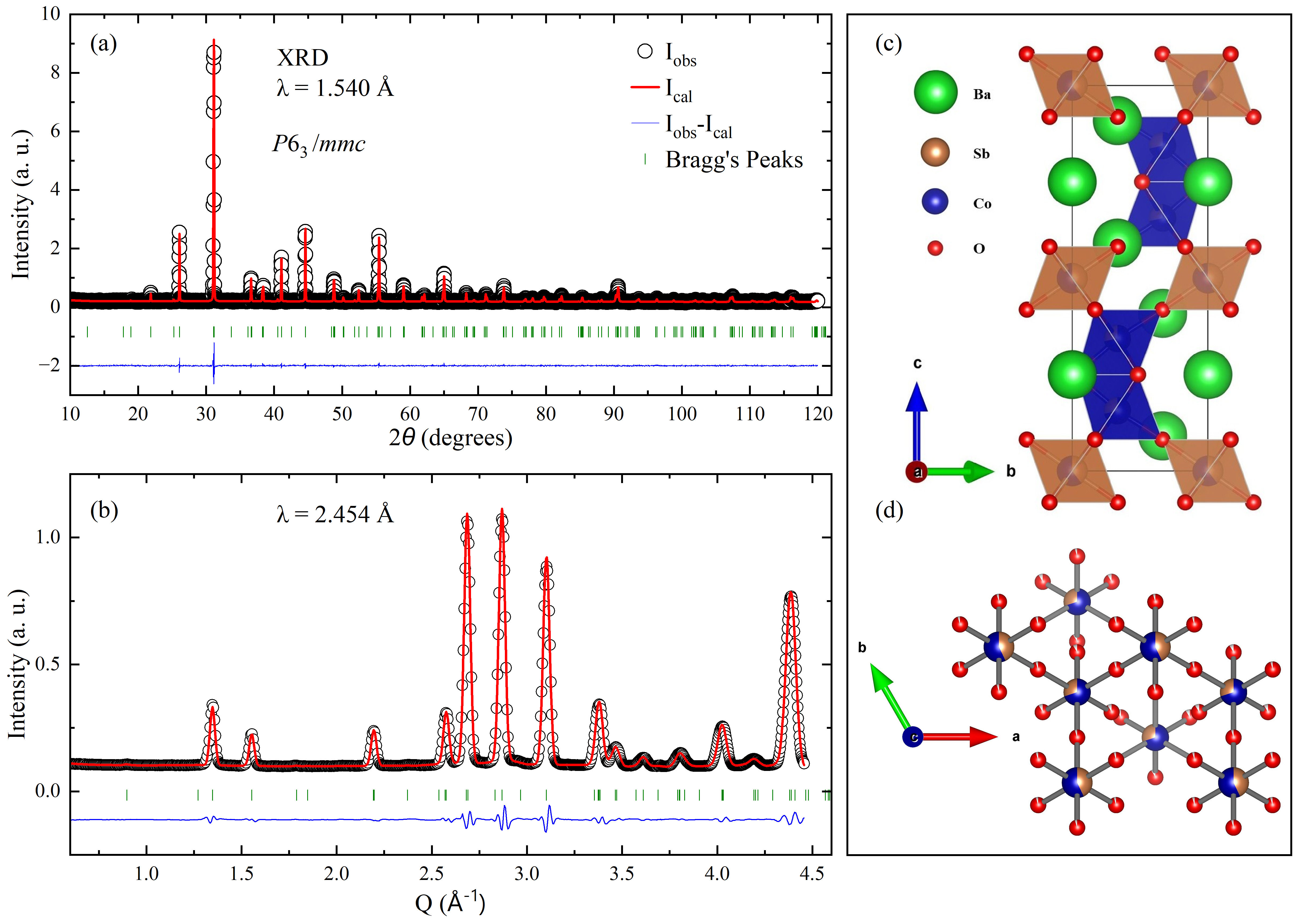}
		\caption{(a) Rietveld refinement of the powder X-ray diffraction data collected at room temperature. (b) Rietveld refinement of the neutron powder diffraction data for \BSCO at 300~K. (c) Refined crystal structure of \BSCO. (d) Co and Sb form a triangular lattice in the ab plane. The Co$_2$O$_9$ dimers, formed by face-sharing CoO$_6$ octahedra, are highlighted in blue along the $c$-axis. The black open circles represent the observed diffraction pattern, the red solid line indicates the calculated fit, and the blue line at the bottom shows the difference between the observed and calculated intensities. The olive vertical markers denote the positions of allowed Bragg reflections.} 
		\label{Fig1}
	\end{figure*}
	
	\subsection{Crystal Structure} 
	
	\subsubsection{X-ray Diffraction}
	
	To determine the crystal structure and assess phase purity, powder X-ray diffraction measurements were performed on \BSCO\ at room temperature. Figure~\ref{Fig1}~(a) displays the XRD pattern along with the Rietveld refinement. The results confirm that \BSCO\ crystallizes in a hexagonal perovskite structure with space group \sg\ (No.~194), consistent with structurally related compounds such as Ba$_3$CoSb$_2$O$_9$ and Ba$_3$NiSb$_2$O$_9$~\cite{Istomin2004, Doi2004, Darie2023}. The refined lattice parameters are $a = 5.7284(2)$~\AA\ and $c = 14.0725(7)$~\AA, yielding a goodness-of-fit (GOF) of 1.03, a profile reliability factor $R_p = 5.31\pct$, and a weighted profile factor $wR_p = 6.68\pct$.
	
	The refined crystal structure is illustrated in Figure~\ref{Fig1}~(c, d). It features Co$_2$O$_9$ dimers formed by face-sharing CoO$_6$ octahedra—structural motifs central to the material’s magnetic behavior. These dimers are linked via corner-sharing SbO$_6$ octahedra, constructing a layered hexagonal framework. In the \textit{ab}-plane, Co and Sb ions generate triangular networks of partially disordered octahedra. The refined site occupancies reveal substantial Co/Sb intersite mixing ($\sim$~30\pct), indicating significant cation disorder that alters the local bonding environment and magnetic exchange pathways.
	
	Earlier studies on Ba$_2$Sb$_{x}$Co$_{2-x}$O$_{6-\delta}$~\cite{Istomin2004} also reported a similar hexagonal structure with notable Sb/Co site disorder. In contrast, our work on \BSCO\ provides a more comprehensive investigation of its thermodynamic and transport behavior, including heat capacity, AC susceptibility, and electrical resistivity. The observed spin-glass-like transition underscores the crucial role of cation disorder and possible mixed valence states in disrupting long-range magnetic order, highlighting the complexity of magnetic interactions in this dimer-based system.

	\subsubsection{Neutron Powder Diffraction}
	
	\begin{table}
		\caption{Refined structural parameters for \BSCO\ (x = 0.04, $\delta$ = 0.54) at 300~K extracted from the Rietveld refinement of neutron powder diffraction data. Space group \sg, No.~194, \(a = 5.7289(2)\)~Å, \(c = 14.0800(12)\)~Å, \(R_p = 2.05\pct\), and \(wR_p = 2.95\pct\).}
		\label{T1}
		\setlength\extrarowheight{4pt}
		\setlength{\tabcolsep}{2pt}
		\begin{tabular}{ccccccc}
			\hline
			Atom & $x$ & $y$ & $z$ & Occ. & U$_{iso}$ & Site \\ \hline
			Ba1 & 0.0000  & 0.0000  & 0.2500  & 1.000 & 0.001(1) & 2$b$ \\ 
			Ba2 & 0.3333 & 0.6667 & 0.0916(5) & 1.000 & 0.008(1) & 4$f$ \\ 
			Co1 & 0.3333 & 0.6667 & 0.84385(6) & 0.696(4) & 0.025(2) & 4$f$ \\ 
			Sb1 & 0.3333 & 0.6667 & 0.84385(6) & 0.304(4) & 0.025(2) & 4$f$ \\ 
			Sb2 & 0.0000 & 0.0000 & 0.0000 & 0.433(2) & 0.001(1) & 2$a$ \\ 
			Co2 & 0.0000 & 0.0000 & 0.0000 & 0.567(2) & 0.001(1) & 2$a$ \\ 
			O1  & 0.5197(6) & 1.0395(13) & 0.25000 & 0.914(1) & 0.010(1) & 6$h$ \\ 
			O2  & 0.8338(4) & 1.6677(8) & 0.0818(1) & 0.952(3) & 0.005(2) & 12$k$ \\ 
			\hline
		\end{tabular}
	\end{table}

	To further investigate the structural and magnetic properties of \BSCO, neutron powder diffraction measurements were performed at 300, 80, 50, 30, 20, and 1.5~K. No additional magnetic Bragg reflections were observed at low temperatures compared to the 300~K pattern, confirming the absence of long-range magnetic ordering down to 1.5~K. Rietveld refinement of the 300~K NPD data (Fig.~\ref{Fig1}(b)) yielded precise atomic positions, site occupancies, and insights into structural disorder.

	To explore the presence of short-range magnetic correlations, we analyzed the low-\(Q\) region of the diffraction patterns at various temperatures. No temperature-dependent diffuse magnetic scattering features were detected, indicating an absence of significant short-range magnetic order. This suggests that any spin correlations within magnetic clusters are confined to very short length scales—too limited to generate observable diffuse features. A similar lack of diffuse scattering has been reported in related cluster spin glass systems such as CaMnFeTaO$_6$~\cite{Kearins2021}, where the extremely short correlation length renders magnetic scattering indistinguishable from the background.
	
	The refined structural parameters (Table~\ref{T1}) reveal two key features: (i) pronounced intersite mixing (~30\pct) between Sb and Co at the 4$f$ Wyckoff position, and (ii) a notable oxygen deficiency (\(\delta = 0.54\)). The cation disorder—arising from the statistical distribution of Co and Sb ions over shared crystallographic sites—generates random local environments, introducing frustration into the magnetic exchange interactions and likely stabilizing the observed spin-glass state. Each unit cell of \BSCO{} contains two crystallographically equivalent Co–Co dimers formed by face-sharing CoO$_6$ octahedra, with an intradimer Co--Co distance of approximately 2.64~\AA{} and a shortest interdimer Co--Co distance of about 5.73~\AA.
	
	Local distortions of the CoO$_6$ octahedra were quantified through bond-angle analysis. For the Co1 site (4$f$), bond angles deviate markedly from the ideal 90\degree, with representative values of 85.1\degree, 92.3\degree, and 88.7\degree. Similar deviations are found at the Co2 site (2$a$), with angles of 84.9\degree, 93.1\degree, and 89.5\degree. Assuming an oxidation state of +5 for Sb, the refined stoichiometry corresponds to an average Co oxidation state of +2.9, consistent with a mixed-valent population of roughly 10\pct\ Co$^{2+}$ and 90\pct\ Co$^{3+}$. This mixed-valence character, coupled with octahedral distortions, generates a locally heterogeneous electronic environment that modulates magnetic exchange. Bond valence sum (BVS) calculations (see Table~\ref{BVS}, Appendix) support this oxidation state distribution.
	
	Oxygen vacancies were refined at both O1 and O2 sites, with a stronger deficiency at the O1 site (\(o(\text{O1}) = 0.914\) vs. \(o(\text{O2}) = 0.952\)). These vacancies predominantly reside within the BaO$_3$ layers that bridge the face-sharing CoO$_6$ dimers, a feature common in 6$H$-hexagonal perovskites such as Ba$_2$ScAlO$_5$~\cite{Murakami2023} and BaFeO$_{2.73}$~\cite{Gomez2001}. The resulting reduction in electrostatic screening between neighboring Co ions can lead to Co--Co distance contraction, as observed in oxygen-deficient BaTiO$_{3-y}$~\cite{Sinclair1999}. Furthermore, the uneven distribution of vacancies promotes the formation of undercoordinated CoO$_4$ tetrahedral motifs, reducing the connectivity of face-sharing octahedra. This structural reconfiguration mirrors features seen in hexagonal analogs of brownmillerite phases~\cite{Calle2009}, where altered coordination and metal-metal separations contribute to enhanced magnetic frustration via disrupted superexchange pathways.

	\subsection{DC Magnetization}
	
	\begin{figure}
		\includegraphics[width=1.0\linewidth]{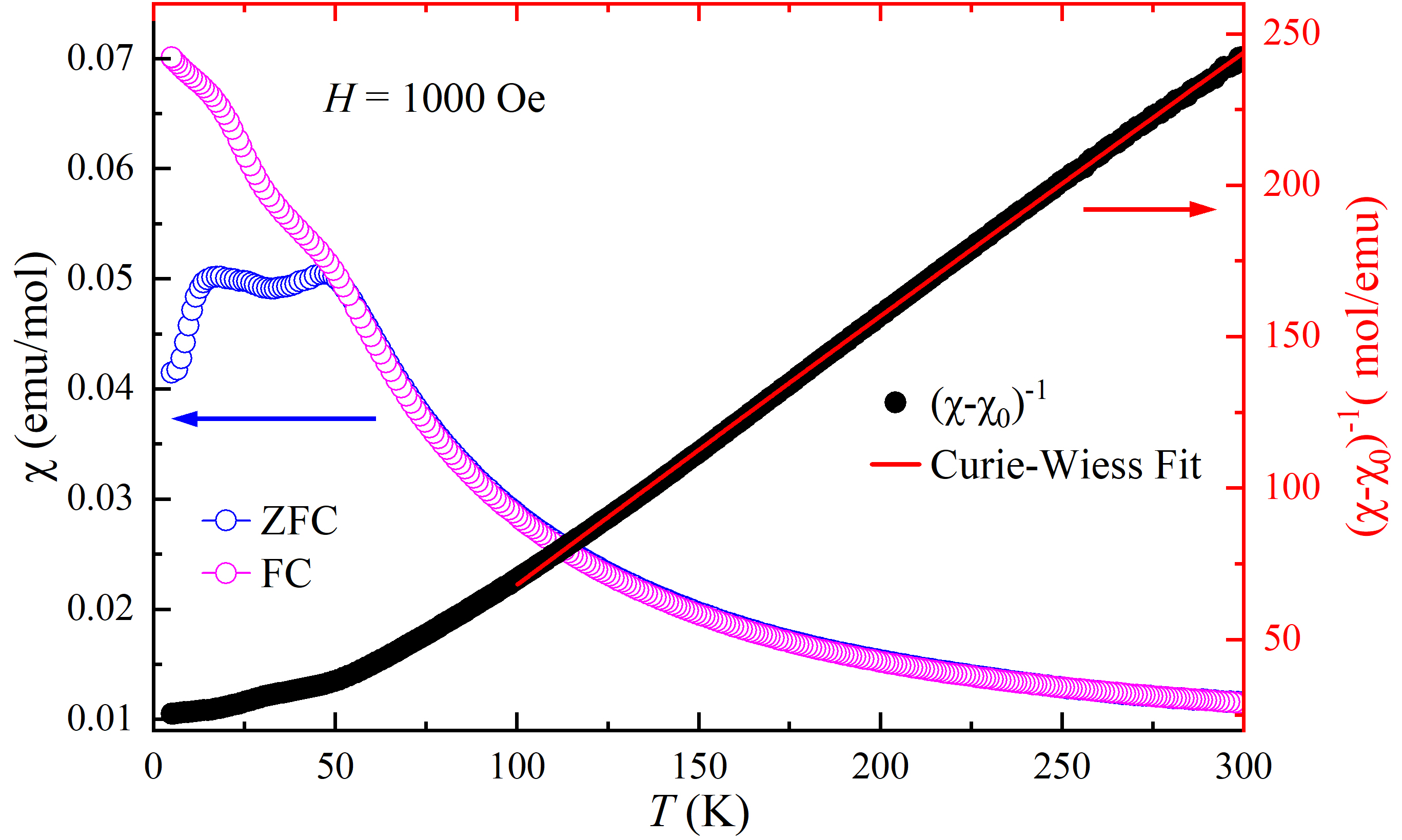}
		\caption{Temperature dependence of magnetic susceptibility (ZFC-FC) (left $y$-axis) and inverse susceptibility with Curie-Weiss fit (right $y$-axis) for \BSCO\ at an applied field of 1000~Oe. The red solid line represents the CW fit in the range 100–300~K.}
		\label{Fig3}
	\end{figure}
	
	Magnetization and susceptibility measurements were performed to investigate the magnetic behavior of \BSCO. Figure~\ref{Fig3}~(a) displays the temperature-dependent magnetic susceptibility measured under zero-field-cooled (ZFC) and field-cooled (FC) conditions, along with the inverse susceptibility, all recorded under an applied magnetic field of 1000 Oe over the temperature range 2–300 K. The left y-axis corresponds to the ZFC-FC susceptibility, while the right y-axis shows the inverse susceptibility. A clear bifurcation between the ZFC and FC curves emerges below approximately 50 K, signaling magnetic irreversibility that may originate from magnetic frustration or a spin-glassy ground state. In the high-temperature region (100–300 K), the susceptibility follows a modified Curie-Weiss (CW) behavior:	
	
	\begin{equation}
		\rchi(T) = \rchi_{0} + \frac{C}{T - \theta},
	\end{equation}
	
	\noindent	where $\rchi_{0}$ is the temperature-independent susceptibility, $C$ is the Curie constant, and $\theta$ is the Curie-Weiss temperature. The effective magnetic moment is determined using $\mu_\textrm{eff} = \sqrt{3k_{B}C/N_{A}}$, where $N_{A}$ is Avogadro’s constant and $k_{B}$ is the Boltzmann constant \cite{Mugiraneza2022}. The fit yields an effective moment of $\mu_{\text{eff}} = 2.96~\mu_B$.
	
	To interpret this effective moment, we consider the formal oxidation states based on stoichiometry. Assuming Ba$^{2+}$, Sb$^{5+}$, and O$^{2-}$, the average oxidation state of Co is calculated as +2.9, suggesting a mixed-valence state with a possible distribution of 10\pct Co$^{2+}$ and 90\pct Co$^{3+}$. To explore how different spin states influence the observed magnetic moment, we consider combinations of high-spin (HS), intermediate-spin (IS), and low-spin (LS) states for both Co$^{2+}$ and Co$^{3+}$. 
	
	The calculated effective magnetic moments for these configurations are summarized in Table~\ref{tab:eff_moment}. The theoretical effective moment for this mixed-valence system is calculated using
	
	\begin{equation}
		\mu_{\text{eff}}^{\text{theory}} = \sqrt{0.1\,\mu_{\text{eff}}(\text{Co}^{2+})^2 + 0.9\,\mu_{\text{eff}}(\text{Co}^{3+})^2}.
	\end{equation}
	
	Although intermediate-spin Co$^{3+}$ is less common, crystal field effects, local lattice distortions, and covalency can stabilize such states in certain cobalt oxides \cite{Korotin1996, Maris2003}.
	
	\begin{table}[h]
		\centering
		\caption{Calculated effective magnetic moments ($\mu_{\text{eff}}$) for various spin state combinations of Co$^{2+}$ (10\pct) and Co$^{3+}$ (90\pct).}
		\label{tab:eff_moment}
		\setlength\extrarowheight{4pt}
		\setlength{\tabcolsep}{6pt}
		\begin{tabular}{ccc}
			\hline
			\textbf{Co$^{2+}$ Spin State} & \textbf{Co$^{3+}$ Spin State} & $\mu_{\text{eff}}$ ($\mu_B$) \\
			\hline
			High-spin (HS) & High-spin (HS) & 4.81 \\
			High-spin (HS) & Intermediate-spin (IS) & 2.95 \\
			High-spin (HS) & Low-spin (LS) & 1.22 \\
			Low-spin (LS) & High-spin (HS) & 4.68 \\
			Low-spin (LS) & Intermediate-spin (IS) & 2.74 \\
			Low-spin (LS) & Low-spin (LS) & 0.55 \\
			\hline
		\end{tabular}
	\end{table}
	
	The experimentally observed $\mu_{\text{eff}} = 2.96~\mu_B$ is closest to the scenario where Co$^{2+}$ adopts a high-spin configuration ($\mu_{\text{eff}} = 3.88~\mu_B$) and Co$^{3+}$ is in an intermediate-spin state ($\mu_{\text{eff}} = 2.8~\mu_B$). Using these values in Equation (1), the resulting theoretical effective moment is $\mu_{\text{eff}}^{\text{theory}} \approx 2.95~\mu_B$ \cite{Amin2024, Chin2017, Oey2020, Korotin1996, Maris2003}, which is in close agreement with the experimental data. 
	
	This supports a mixed-valence state comprising Co$^{2+}$ and Co$^{3+}$, consistent with an average oxidation state of +2.9. While this scenario is plausible, the stabilization of an intermediate-spin state for Co$^{3+}$ remains speculative and warrants further spectroscopic or structural studies.
	
	The mixed-valent state, combined with oxygen vacancies primarily located at the oxygen sites, likely plays a significant role in modifying local magnetic interactions and electronic properties \cite{Podlesnyak2008, Burnus2008}. Furthermore, the extracted Curie–Weiss temperature of $\theta \sim 21$~K is positive, indicative of dominant ferromagnetic interactions and suggesting that the magnetic behavior is primarily governed by interactions at the dimer level rather than between individual Co spins. In \BSCO, the face-sharing CoO$_6$ dimers likely host antiferromagnetic intradimer exchange interactions, facilitated by direct Co--Co orbital overlap. However, ferromagnetic correlations between these dimers within the triangular \textit{ab} plane may arise via longer-range superexchange pathways through bridging oxygens or via polarizable Sb–O units. This combination of AFM intradimer and FM interdimer couplings can collectively yield a net positive Curie–Weiss temperature. A similar scenario has been discussed in the trimer-based system Ba$_4$NbMn$_3$O$_{12}$, where antiferromagnetic interactions dominate within Mn$_3$ clusters, but ferromagnetic coupling between clusters leads to overall positive $\theta$ \cite{Streltsov2018}.
	
	A noteworthy feature is the positive value of $\rchi_{0} = 2.7 \times 10^{-3}$ emu/mol, which is suggestive of Van Vleck paramagnetism \cite{Mugiraneza2022}. This observation is consistent with a scenario where local moments remain partially aligned even at elevated temperatures, likely due to remnant interactions within the Co dimers. Similar behavior has been observed in structurally related systems such as Ba$_4$NbMn$_3$O$_{12}$~\cite{Nguyen2019}, Ba$_4$TaMn$_3$O$_{12}$~\cite{Ali2024}, Ba$_{4}$NbRu$_{3}$O$_{12}$~\cite{Nguyen2018}, and Ba$_{4}$Nb$_{0.8}$Ir$_{3.2}$O$_{12}$~\cite{Thakur2020}.
	
	\begin{figure}
		\includegraphics[width=1.0\linewidth]{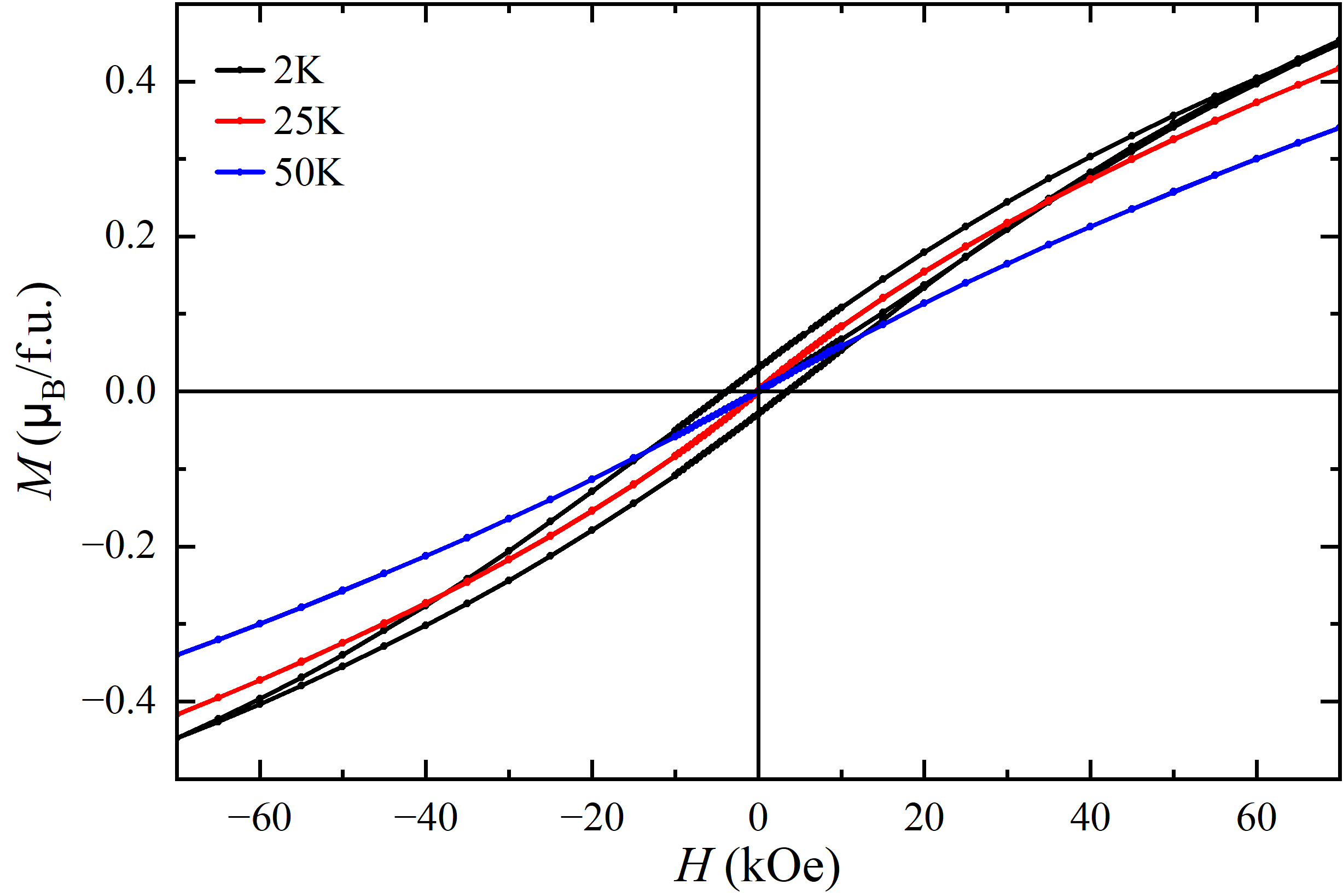}
		\caption{Isothermal magnetization ($M$ vs $H$) curves at different temperatures ($T$ = 2, 25, and 50~K).}
		\label{Fig4}
	\end{figure}
	
	To further explore the magnetic behavior, isothermal magnetization ($M$ vs $H$) measurements were conducted at 2, 25, and 50~K, as shown in Figure~\ref{Fig4}. The magnetization curves show a gradual increase with applied field, exhibiting slight curvature at low fields and approaching linear behavior at higher fields. At 2~K, a weak hysteresis loop is observed, indicative of minor magnetic irreversibility. Notably, the magnetization does not saturate up to 70~kOe, reaching a maximum value of only $\sim 0.45~\mu_B$ per formula unit. This is significantly lower than the expected theoretical saturation moment of $M_{\text{sat}} \sim 2.1~\mu_B$ per formula unit for a mixed-valence Co$^{2+}$ (high spin) and Co$^{3+}$ (intermediate spin) scenario. The absence of sharp metamagnetic transitions or magnetization plateaus further suggests a lack of long-range magnetic order, consistent with a magnetically frustrated state.
	
	The strongly reduced magnetization and absence of saturation indicate the presence of persistent spin dynamics and frozen short-range correlations, characteristics often associated with spin-glass-like behavior. The weak hysteresis observed at 2~K further supports the presence of glassy magnetic dynamics. Such behavior typically arises from competing interactions, local disorder, and geometric frustration within the Co dimers, which collectively hinder the establishment of long-range magnetic order.
	
	The coexistence of Co$^{2+}$ and Co$^{3+}$ in a mixed-valence state, along with possible oxygen vacancies and structural distortions, likely enhances the degree of magnetic frustration. These factors may promote collective spin freezing, similar to phenomena observed in related cobalt oxides \cite{Podlesnyak2008, Burnus2008, Shibasaki2011}.

	\subsection{Heat Capacity}
	
	\begin{figure*}
		\includegraphics[width=1.0\linewidth]{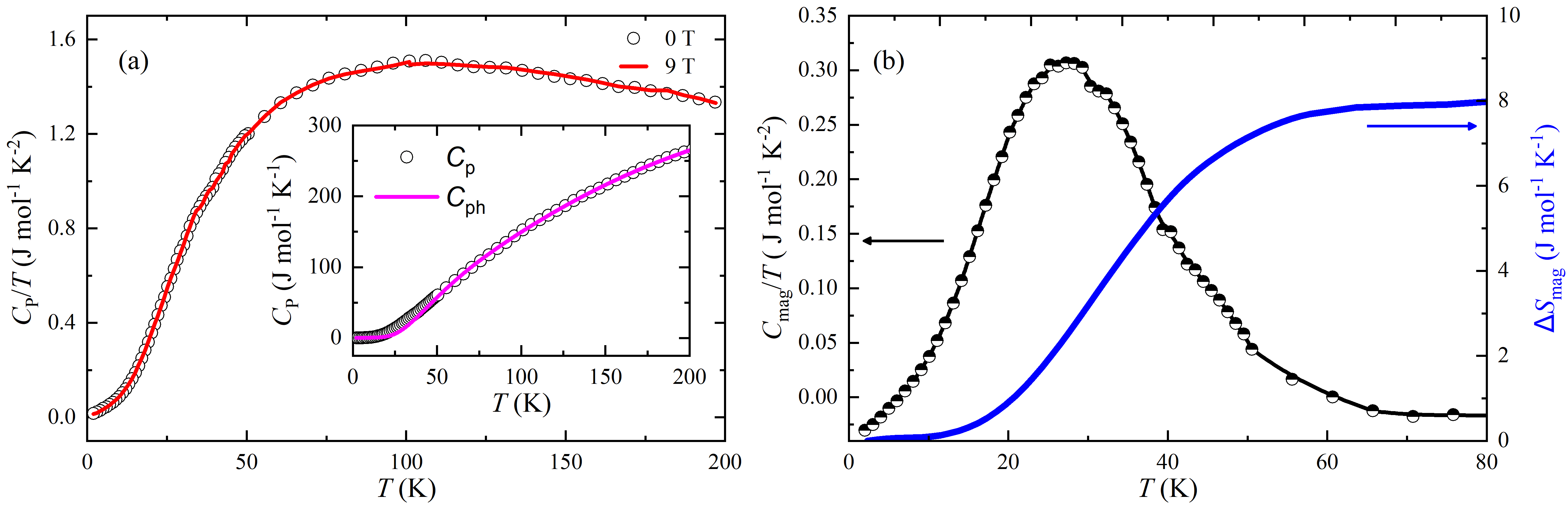}
		\caption{Heat capacity ($C_{\rm p}$) data analysis. (a) Temperature dependence of $C_{\rm p}/T$ measured at zero and 9~T between 2 and 200~K. The inset shows the total measured heat capacity (symbols) overlaid with the estimated phononic contribution at zero field (magenta line; see text). (b) Magnetic contribution to heat capacity ($C_{\rm p}$-$C_{\rm ph}$) divided by temperature (left $y$-axis) and the evolution of the magnetic entropy, $\Delta S_{\rm mag}(T)$ (right $y$-axis).}
		\label{Fig5}
	\end{figure*}
	
	To further investigate the anomalous magnetic behavior observed in the susceptibility measurements, we performed heat capacity measurements on the same sample. Figure~\ref{Fig5}(a) presents the specific heat divided by temperature ($C_{\rm p}/T$) between 2 and 200~K at 0 and 9~T. A broad feature is observed around 30~K, indicative of short-range magnetic correlations. Notably, applying a 9~T field suppresses this feature at lower temperatures, confirming its magnetic origin.
	
	In insulating magnetic materials, the total heat capacity ($C_{\rm p}$) consists of contributions from phononic ($C_{\rm ph}$) and magnetic ($C_{\rm mag}$) degrees of freedom. Ideally, these components can be separated by measuring a nonmagnetic structural analog; however, without such a reference compound, we employed a Debye-Einstein model~\cite{Kittel2004, Sebastian2021} to approximate the lattice contribution. The phonon heat capacity was modeled using a combination of one Debye and one Einstein term, given by: 
	\begin{equation}
		C_{\rm ph} (T) = f_{D}C_{D}(\theta_{D}, T) +  g_{E}C_{E}(\theta_{E}, T),
		\label{D_E}
	\end{equation}
	
	\noindent	where the Debye contribution, representing acoustic phonon modes, is expressed as:
	
	\begin{equation}
		C_{D}(\theta_{D}, T) = 9nR\left({\frac{T}{\theta_{D}}}\right)^{3} \int_{0}^{\theta_{D}/T} \frac{x^4e^x}{(e^x-1)^2} dx,
		\label{Debye}
	\end{equation}
	
	\noindent with $n$ being the number of atoms per formula unit, $R$ the universal gas constant, $\theta_{D}$ the Debye temperature, and $x = \frac{\hbar \omega}{k_{B}T}$. The Einstein term, accounting for optical phonon modes, follows:
	
	\begin{equation}
		C_{E}(\theta_{E}, T) = 3nR\left( \frac{\theta_{E}}{T} \right)^2 \frac{e^{\theta_{E}/T}}{(e^{\theta_{E}/T}-1)^2},
		\label{Eienstien}
	\end{equation}
	
	\noindent	where $\theta_{E}$ is the Einstein temperature. Given that \BSCO\ has 15 atoms per formula unit, the sum of $f_{D}$ and $g_{E}$ is constrained to 15, ensuring the model accounts for all lattice vibrations.  
	
	The best fit of $C_{\rm p}$ data between 50 and 200~K yields $\theta_{D} \approx 776$~K and $\theta_{E} \approx 190$~K, with $f_{D} \approx 9$ and $g_{E} \approx 6$ [magenta solid line in the inset of Fig.~\ref{Fig5}(a)]. Alternative models incorporating additional Einstein terms or a purely Debye-based fit yielded a poorer agreement, justifying our minimal one-Debye, one-Einstein model. By subtracting the fitted phononic contribution from the measured $C_{\rm p}$ at zero field, we extract the magnetic contribution, $C_{\rm mag}$, which is plotted in Fig.~\ref{Fig5}(b).
	
	The temperature dependence of $C_{\rm mag}/T$ reveals a broad hump centered around 30~K, a hallmark of short-range spin correlations rather than a sharp $\lambda$-type anomaly associated with long-range magnetic ordering. This behavior is consistent with a spin-glass or cluster-glass state, where magnetic moments freeze gradually rather than undergoing a sharp phase transition. The suppression of this feature under a 9~T field further supports this interpretation, suggesting that applied fields destabilize frozen spin configurations.
	
	Figure~\ref{Fig5}(b) also shows the evolution of the magnetic entropy ($\Delta S_{\rm mag}$) as a function of temperature (blue solid line). The total magnetic entropy change is obtained by integrating the magnetic heat capacity ($C_{\rm mag}$) divided by temperature over the 2–80~K range:
	
	\begin{equation}
		\Delta S_{\rm mag}(T) = \int_{2~\text{K}}^{80~\text{K}} \frac{C_{\rm mag}}{T} dT.
	\end{equation}
	
	\noindent This integration yields a total entropy change of $\Delta S_{\rm mag} = 8.15$~J~mol$^{-1}$~K$^{-1}$. For the mixed-valence state of Co$^{2+}$ (high spin, $S = 3/2$) and Co$^{3+}$ (intermediate spin, $S = 1$), the theoretical magnetic entropy is calculated as  
	$\Delta S_{\rm mag} = 0.1 \cdot R \ln(4) + 0.9 \cdot R \ln(3) \approx 9.38~\text{J~mol}^{-1}~\text{K}^{-1}.$  The observed entropy change is reasonably close to this theoretical value, with the slight deficit possibly arising from residual quantum fluctuations, partial spin freezing, or covalency effects that suppress the full recovery of spin entropy. Such behavior is often observed in frustrated systems and cluster spin glasses, where competing interactions and local structural distortions hinder complete spin alignment even at low temperatures.
	
	Overall, the heat capacity data reinforce the presence of a magnetically frustrated state in \BSCO, characterized by broad spin freezing and persistent short-range correlations.
	
	\subsection{AC Susceptibility}
	
	\begin{figure}
		\includegraphics[width=1.0\linewidth]{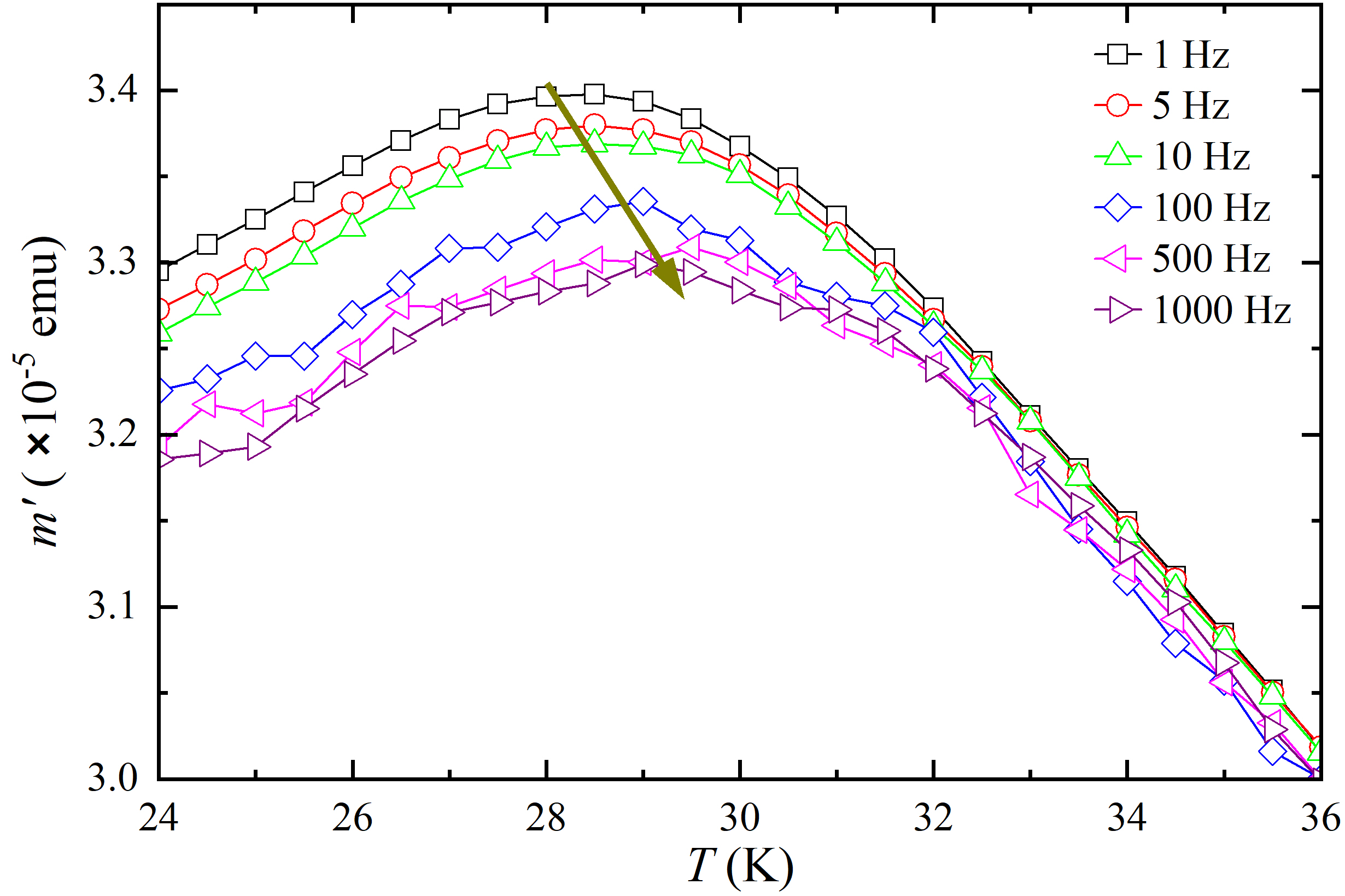}
		\caption{Temperature dependence of the real part of the AC magnetization measured at various frequencies.}
		\label{Fig6}
	\end{figure}
	
	To investigate the dynamical nature of the spin-glass state in \BSCO, we performed AC susceptibility measurements as a function of frequency (1–1000~Hz) and temperature (2–50~K). The sample was cooled to 2~K before applying a small AC magnetic field of 10~Oe to assess its temperature-dependent response. The real part of the AC susceptibility, recorded between 24~K and 36~K, is presented in Fig.~\ref{Fig6}. A broad peak appears near 28.3~K for \(\nu = 1\)~Hz, and this peak systematically shifts to higher temperatures with increasing frequency. The peak positions determined for various frequencies are listed in Table~\ref{T2}. These temperatures were extracted from the minima of the first susceptibility derivative with respect to temperature. The frequency-dependent shift in peak position indicates a delay in spin relaxation, characteristic of a glassy magnetic state. Additionally, the gradual reduction in peak height with increasing frequency is a well-known signature of spin-glass behavior.
	
	\begin{table}
		\centering
		\begin{tabular}{cc}
			\toprule
			Frequency (Hz) & Peak Temperature (K) \\
			\midrule
			5     & 28.4 \\
			10    & 28.5 \\
			100   & 29.0 \\
			500   & 29.4 \\
			1000  & 29.5 \\
			\bottomrule
		\end{tabular}
		\caption{Peak temperature as a function of frequency.}
		\label{T2}
	\end{table}
	
	To quantify the frequency dependence of the freezing temperature, we employed the Mydosh parameter \(S\), which characterizes the relative shift in peak position with frequency and serves as a diagnostic tool for distinguishing different types of spin-glass systems \cite{Mydosh1982, Mydosh1996}:
	
	\begin{equation}
		S = \frac{\Delta T_f}{T_f \Delta \log_{10}(\nu)},
		\label{eq:S}
	\end{equation}
	
	\noindent	where \(\Delta \log_{10}(\nu) = \log_{10}(\nu_2) - \log_{10}(\nu_1)\) and \(\Delta T_f = (T_f)_{\nu_2} - (T_f)_{\nu_1}\). For canonical spin glasses, \(S\) typically falls within the range of 0.005–0.01, whereas in cluster-glass systems, it varies between 0.01 and 0.05~\cite{Mydosh1993}. Our analysis yields an average \(S\) value of 0.016 $\pm$ 0.003, consistent with a cluster-glass system \cite{Bag2018}.
	
	\begin{figure}
		\includegraphics[width=1.0\linewidth]{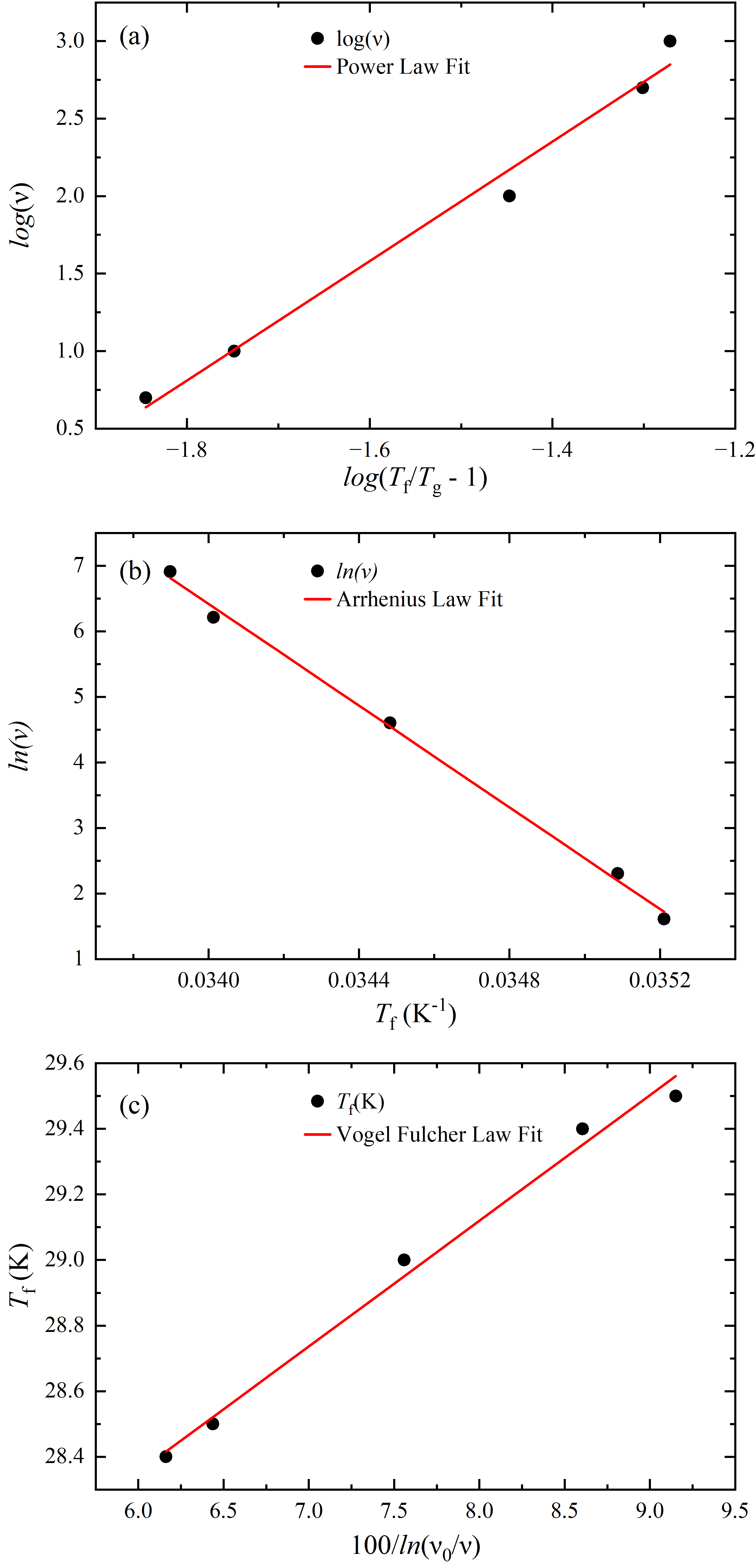}
		\caption{Fitting of the frequency dependence of the freezing temperature using (a) the power law, (b) the Arrhenius law, and (c) the Vogel-Fulcher law.}
		\label{Fig7}
	\end{figure}
	
	We applied different theoretical models to the AC susceptibility data to further analyze the glassy dynamics. The critical slowing down of spin relaxation can be described using the power-law scaling equation:\cite{Binder1986, Mydosh1993, Lago2012}.
	
	\begin{equation}
		\tau = \tau_0 \left( \frac{T_f - T_g}{T_g} \right)^{-z\nu'},
		\label{eq:tau}
	\end{equation}
	
	\noindent	where \(\tau\) represents the characteristic relaxation time corresponding to the frequency \(\nu\), \(T_g\) is the freezing temperature in the \(\nu \to 0\) limit, \(\tau_0\) is the spin-flip time of an individual spin or cluster, and \(z\nu'\) is the critical exponent related to the correlation length. This equation is often rewritten as:
	
	\begin{equation}
		\log_{10} \nu = \log_{10} \nu_0 + z\nu' \log_{10} \left( \frac{T_f - T_g}{T_g} \right),
		\label{eq:log_nu}
	\end{equation}
	
	\noindent	where \(\nu = 2\pi/\tau\). As depicted in Fig.~\ref{Fig7}(a), the plot of \(\log_{10} \nu\) vs \(\log_{10} (T_f/T_g -1)\) follows a linear trend. With \(T_g\) fixed at 28 K, the power-law fit yields \(z\nu' = 3.95\) and \(\tau_0 = 1.12 \times 10^{-7}\) s. These values are consistent with a cluster-glass system, as canonical spin glasses typically exhibit \(z\nu'\) in the range of 10–12 and \(\tau_0\) around \(10^{-12}\)–\(10^{-13}\) s, whereas cluster glasses generally show \(z\nu'\) between 4 and 6 and \(\tau_0\) in the range of \(10^{-7}\)–\(10^{-9}\) s~\cite{Souletie1985, Shtrikman1981}.
	
	We also examined the Arrhenius law, which describes systems with weakly interacting spins:\cite{Binder1986}.
	
	\begin{equation}
		\tau = \tau^* \exp \left( \frac{E_a}{k_B T_f} \right),
		\label{eq:arrhenius}
	\end{equation}
	
	\noindent	where \(E_a/k_B\) represents the activation energy of the relaxation barriers. The fit, shown in Fig.~\ref{Fig7}(b), results in an unphysically small \(\tau^* = 5.12 \times 10^{-60}\) s and an unrealistically large \(E_a/k_B = 3880.57\) K. These findings confirm that the Arrhenius model is unsuitable for our system, likely due to the presence of interacting magnetic clusters rather than isolated spin entities.
	
	Finally, we applied the Vogel-Fulcher law, which accounts for interactions among clusters:\cite{Mydosh1993, Shtrikman1981}.
	
	\begin{equation}
		\tau = \tau_0 \exp \left( \frac{E_a}{k_B (T_f - T_0)} \right),
		\label{eq:vogel-fulcher}
	\end{equation}
	
	\noindent	where \(T_0\) represents the interaction strength among magnetic entities. The corresponding linear fit in Fig.~\ref{Fig7}(c) yields \(E_a/k_B = 38.30\) K and \(T_0 = 26.05\) K. The nonzero \(T_0\) indicates the presence of interacting clusters, further supporting the cluster-glass interpretation.
	
	The interaction strength can also be assessed using the Tholence criterion~\cite{Souletie1985, Beauvillain1986}:
	
	\begin{equation}
		\delta T_{Th} = \frac{T_f - T_0}{T_0}.
		\label{eq:tholence}
	\end{equation}
	
	For \BSCO, we obtain \(\delta T_{Th} = 0.11\), consistent with values reported for other cluster-glass systems \cite{Mathieu2010,Rahul2021}. These findings collectively establish \BSCO as a cluster glass with significant antisite disorder, distinguishing it from conventional spin-glass systems.
	
	\subsection{Estimation of Cluster Size}
	
	\begin{figure}
		\includegraphics[width=0.95\linewidth]{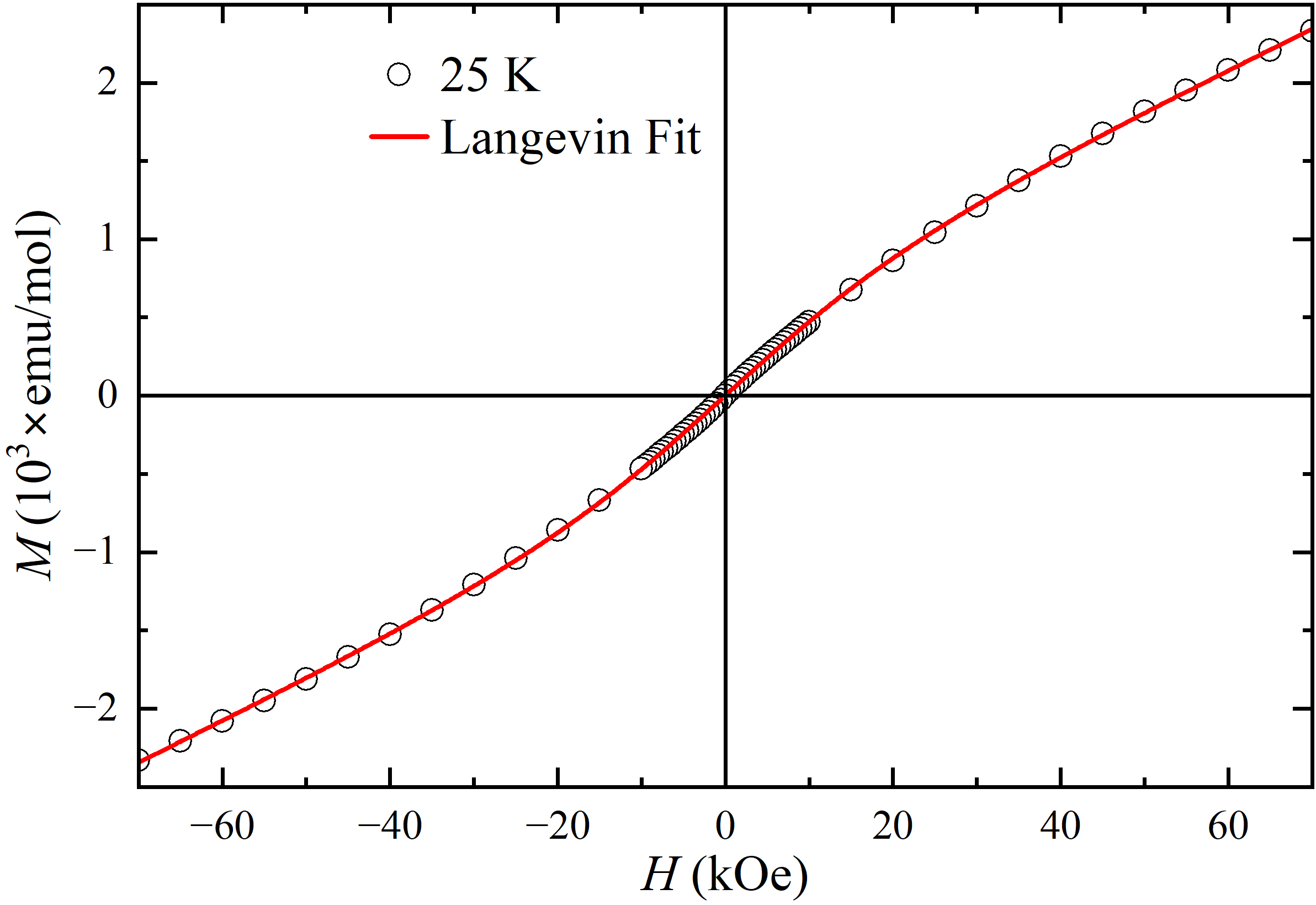}
		\caption{Estimation of the cluster size using the Langevin function.}
		\label{Fig8}
	\end{figure}
	
	The characteristic size of magnetic clusters in \BSCO\ can be estimated by fitting the isothermal magnetization data to the Langevin function, which describes the behavior of paramagnetic or superparamagnetic entities. The Langevin equation is given by~\cite{McCabe1997}:
	
	\begin{equation}
		M(H) = N \mu \left[ \coth \left( \frac{\mu H}{k_B T} \right) - \frac{k_B T}{\mu H} \right] + \rchi H,
		\label{eq:langevin}
	\end{equation}
	
	\noindent	where \( N \) is the number of magnetic clusters per mole, \( \mu \) is the magnetic moment per cluster (in \(\mu_B\)), and \( \rchi \) represents a field-independent Curie-type contribution arising from noninteracting spins~\cite{Anil2020}. At low magnetic fields, the first term dominates due to cluster magnetization saturation, whereas at high fields, the linear \(\rchi H\) term becomes significant. The isothermal magnetization data at 25~K were fitted using Eq.~\eqref{eq:langevin}, as shown in Fig.~\ref{Fig8}.
	
	To estimate the cluster size, we first determine \( N \) from the fitting parameters \( N\mu \) and \( \mu / k_B T \). Assuming a spherical cluster geometry, the cluster volume can be expressed as $V = \frac{4}{3} \pi R^3$, where \( R \) is the average cluster radius. Additionally, the volume can be related to the sample density \( \rho \) as $V = M/\rho$.
	
	Equating these two expressions, the cluster radius is given by
	
	\begin{equation}
		R = \left( \frac{3M}{4\pi \rho N} \right)^{1/3}.
	\end{equation}
	
	From the fitted parameters, we obtain \( N = 19.0235 \times 10^{24} \), leading to an estimated average cluster size of \( R \approx 1.68 \) nm. This value is consistent with those reported for other cluster-glass systems, further supporting the presence of magnetically interacting clusters in \BSCO~\cite{Arun2019}.

	\subsection{Non-equilibrium Dynamics}
	
	\begin{figure}
		\includegraphics[width=1.0\linewidth]{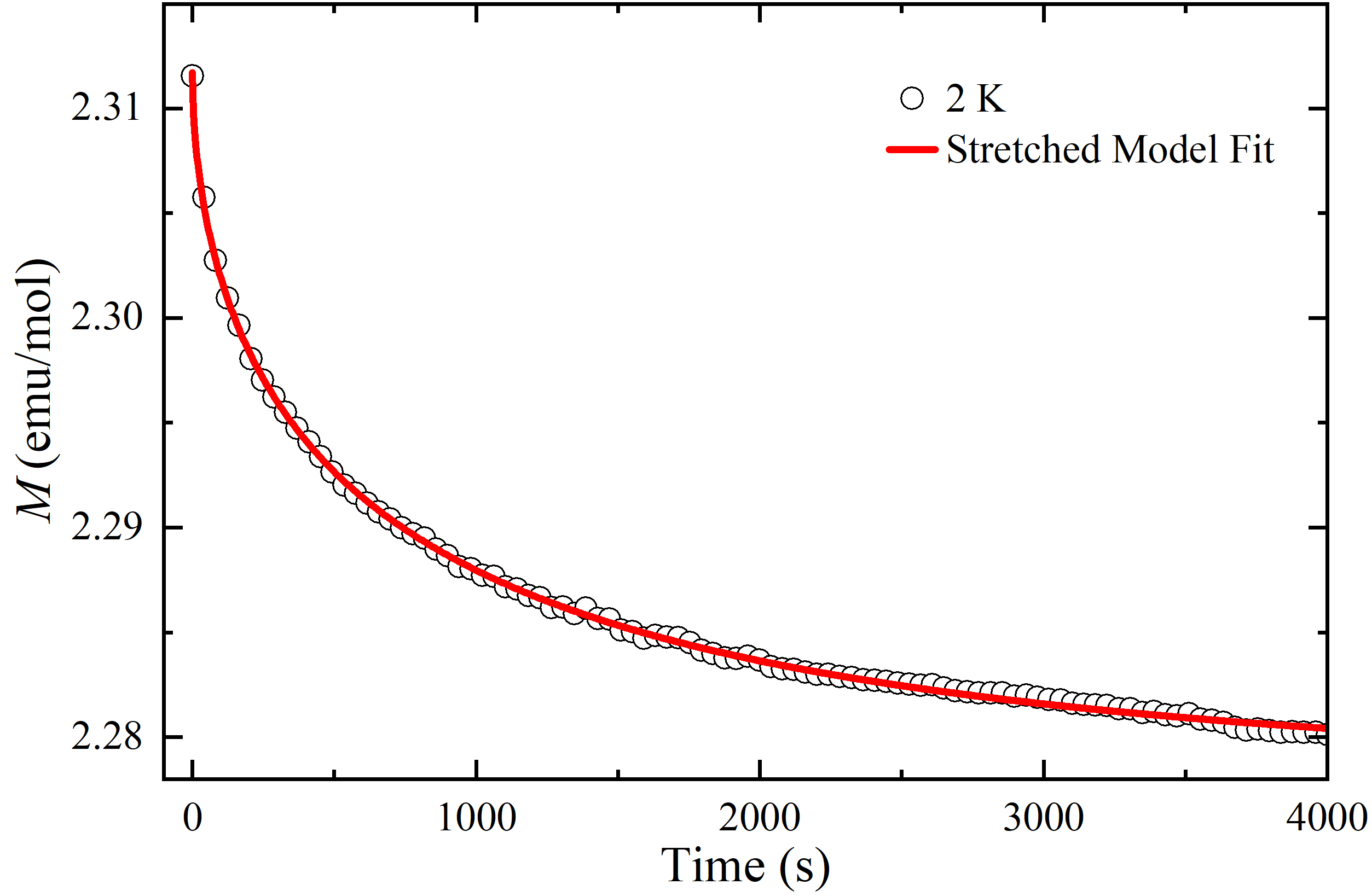}
		\caption{Time-dependent relaxation of field-cooled magnetization after a waiting time of $t = 100$~s.}
		\label{Fig9}
	\end{figure}
	
	\subsubsection{Thermoremanent Magnetization}
	
	Observing magnetic relaxation below the freezing temperature is another hallmark of spin-glass behavior~\cite{Jonason1998, Mathieu2010}. To investigate the thermoremanent magnetization, the sample was cooled to a temperature well below the spin-glass transition temperature following the field-cooled protocol. The sample was cooled to 2~K in a weak applied magnetic field of 100~Oe. Upon reaching 2~K, the field was removed, and after a waiting time of 100~s, the time evolution of magnetization was recorded for 4,000~s, as shown in Fig.~\ref{Fig9}. The magnetization exhibits a characteristic non-equilibrium relaxation, following an exponential decay with time, a signature of spin-glass dynamics.
	
	The time-dependent magnetization data were analyzed using the well-established stretched exponential model~\cite{Freitas2001}, given by:
	
	\begin{equation}
		m(t) = m_0 - m_g \exp \left[ -\left( \frac{t}{\tau} \right)^\beta \right],
	\end{equation}
	
	\noindent	where \( m_0 \) represents the static component of the magnetization, \( m_g \) is the glassy magnetization component, \( \tau \) is the mean relaxation time, and \( \beta \) is the stretching exponent. Although this model lacks a rigorous theoretical foundation, it has been widely used to characterize magnetic relaxation phenomena in spin-glass systems~\cite{Chamberlin1984}. 
	
	For a spin glass, \( \beta \) typically lies between 0 and 1. A value of \( \beta = 0 \) implies no relaxation, whereas \( \beta = 1 \) corresponds to a purely exponential relaxation process~\cite{Mydosh1993}. The stretching exponent thus provides insight into the distribution of relaxation times within the system. The best-fit parameters obtained from the analysis are: \( m_0 = 2.277 \) emu/mol or \(6.957 \times 10^{-5}\) emu, \( \tau = 704 \) s, and \( \beta = 0.548 \). The nonzero values of these parameters confirm the presence of a magnetically frustrated phase~\cite{Rahul2021, Cardoso2003,  Chu1994}. Furthermore, the stretched exponential behavior suggests that the system evolves through a complex energy landscape consisting of multiple metastable states~\cite{Bhattacharyya2011, Pakhira2016}.  
	
	\subsubsection{Magnetic Memory Effect}
	
	\begin{figure}
		\includegraphics[width=1.0\linewidth]{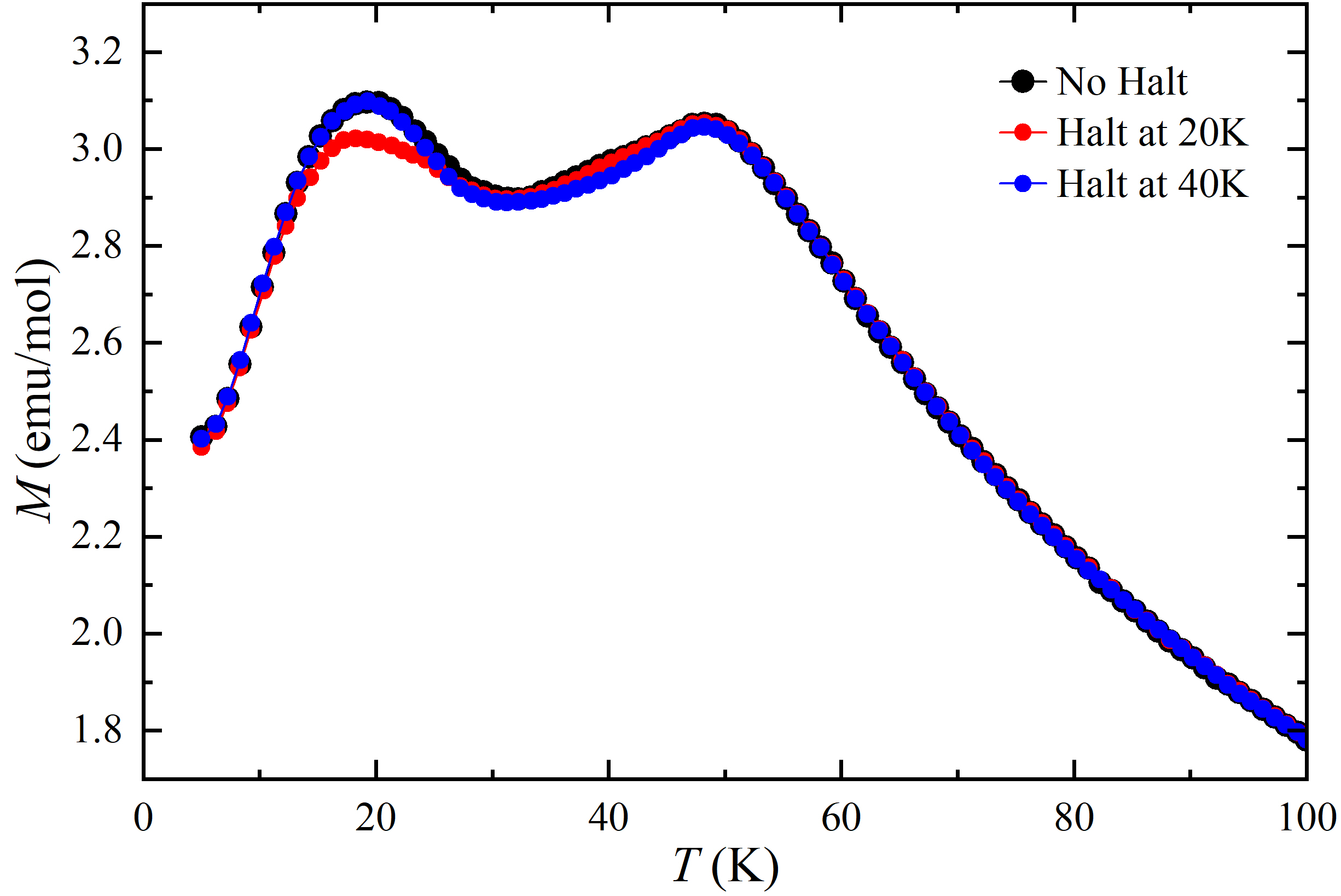}
		\caption{DC magnetization measured with and without halts at two different temperatures ($T$ = 20, and 40~K).}
		\label{Fig10}
	\end{figure}
	
	To probe the low-temperature magnetic dynamics further, we performed a magnetic memory measurement using the zero-field-cooled protocol~\cite{Jonason1998}. In this experiment, the sample was first cooled from room temperature to the target temperature at a constant rate of 3~K/min and then held (halted) at that temperature for 2 hours. Following this dwell period, the sample was cooled further to 2~K. Magnetization was then recorded during the warming cycle under an applied magnetic field of 100~Oe. For comparison, a reference measurement (referred to as the "no-halt" curve) was carried out by cooling the sample continuously from room temperature down to 2~K at the same rate without any intermediate halts.
	
	Two halt temperatures were selected: 40~K and 20~K. The results are shown in Fig.~\ref{Fig10}. For the halt at 40~K, the magnetization curve displays a weak shoulder near the halt temperature, while for the 20~K halt, a distinct dip is observed at 20~K in the warming curve. These features indicate that the system retains a memory of the thermal history, suggesting the presence of a metastable glassy magnetic state below the irreversible temperature.

	The observed memory effect can be understood within the framework of spin-glass physics, where the energy landscape is characterized by many valleys separated by barriers of varying heights~\cite{Mamiya2015}. At a given temperature \( T_i \), relaxation processes are governed by thermally activated transitions over energy barriers comparable to the thermal energy \( k_B T_i \). While higher-energy barriers remain inaccessible, relaxation through lower-energy barriers proceeds efficiently, leading to a redistribution of the system into deeper energy minima. 
	
	Upon further cooling, relaxation mechanisms at lower-energy scales dominate, largely independent of the previous equilibration at \( T_i \). However, when the system is reheated, the previously equilibrated energy barriers at \( k_B T_i \) become reactivated, leading to a slowdown in the relaxation process. This manifests as memory dips in the magnetization curve, confirming that the system retains a record of its thermal history. Such behavior is a defining characteristic of spin glasses, highlighting the presence of a rugged energy landscape with a hierarchy of metastable states.
	
	\subsection{Electrical Resistivity}
	
	\begin{figure}
		\includegraphics[width=1.0\linewidth]{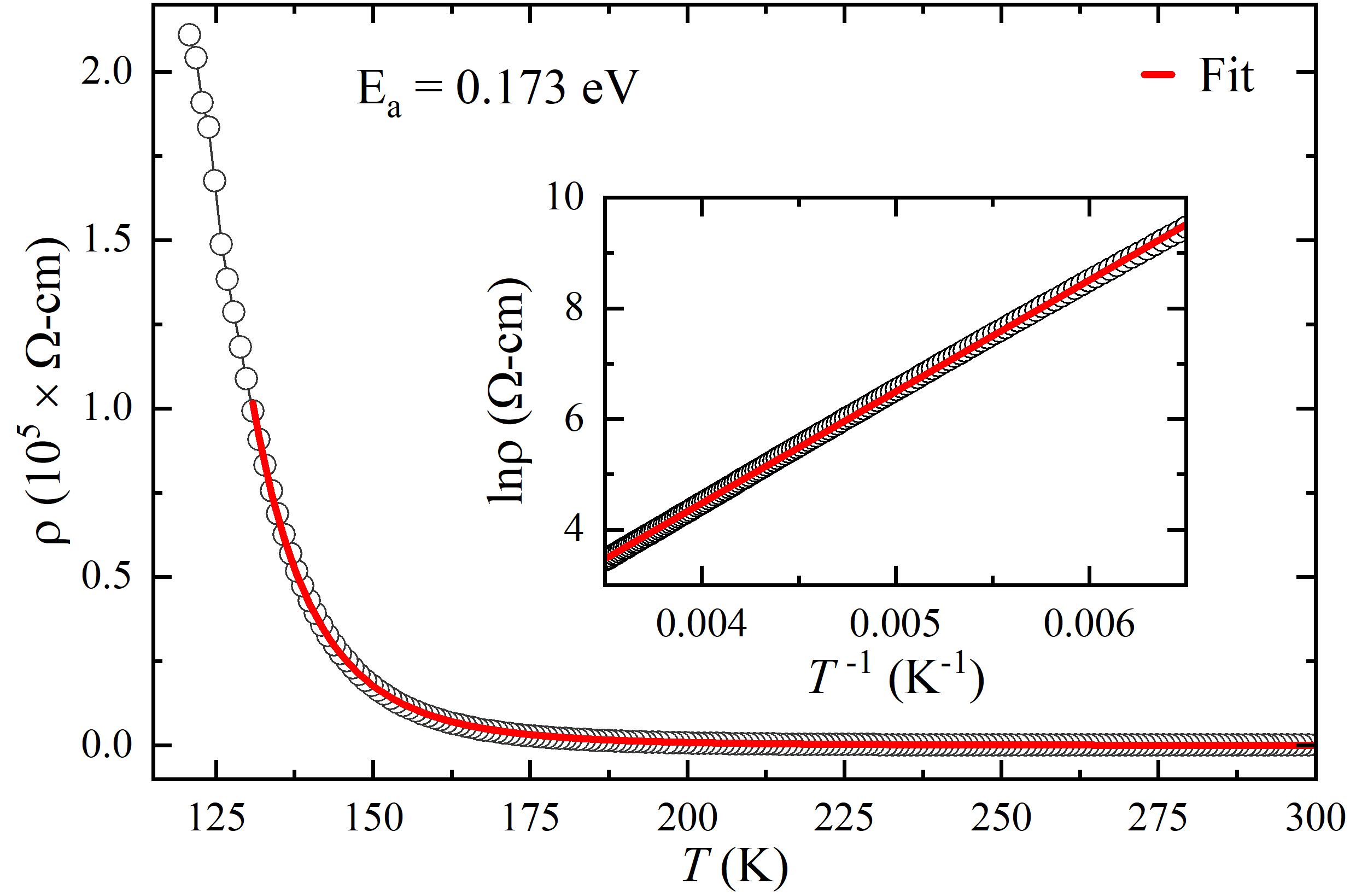}
		\caption{Temperature-dependent DC resistivity of \BSCO in zero applied magnetic field. The main panel shows the resistivity (\(\rho\)) plotted as a function of temperature (120–300~K), while the inset presents the Arrhenius fit in the form of \(\ln\rho\) vs. \(T^{-1}\).}
		\label{Fig11}
	\end{figure}
	
	Figure~\ref{Fig11} presents the temperature-dependent resistivity (\(\rho\)) measured in zero magnetic fields from 120 to 300~K. Below 120~K, the resistivity exceeds the measurable range of our PPMS system, indicating a highly insulating behavior at low temperatures. The resistivity data in the measured temperature range exhibit an activated behavior, well-described by the Arrhenius relation:
	
	\begin{equation}
		\rho(T) = \rho_0 \exp \left( \frac{E_a}{k_B T} \right),
	\end{equation}
	
	\noindent	where \(E_a\) is the activation energy, \(\rho_0\) is a prefactor, and \(k_B\) is the Boltzmann constant. A linear fit to the \(\ln \rho\) vs. \(T^{-1}\) plot (inset of Fig.~\ref{Fig11}) yields \(E_a = 1791 \pm 4\)~K (\(\sim 0.173\)~eV) and \(\rho_0 = 0.115 \pm 0.004\ \Omega\cdot\text{cm}\), confirming thermally activated transport.
	
	The observed activated transport behavior suggests that \BSCO is a band insulator or a polaronic semiconductor. The extracted activation energy (\(\sim 0.173\)~eV) is comparable to the band gap of narrow-gap semiconductors or the small-polaron hopping energy observed in transition-metal oxides \cite{Yuan2019, Liu2014}. The absence of a successful fit with the Mott variable-range hopping (VRH) model~\cite{Hill1976} implies that the conduction mechanism is not dominated by hopping between localized states near the Fermi level, ruling out strong disorder-induced localization.
	
	The large resistivity at lower temperatures and the failure of the VRH model suggest that charge transport is likely governed by intrinsic band-like conduction rather than disorder-driven hopping. This behavior is consistent with insulating oxides containing Co$^{3+}$, where the electronic gap arises from the crystal-field splitting of \(d\)-orbitals and the strong on-site Coulomb interaction (\(U\))~\cite{Chen2011}. Face-sharing CoO$_6$ octahedral dimers in \BSCO may also contribute to the insulating nature, as direct Co–Co interactions could lead to an electronic band structure with a significant gap.
	
	\subsection{Temperature-Dependent X-ray Diffraction}
	
	\begin{figure}
		\includegraphics[width=1.0\linewidth]{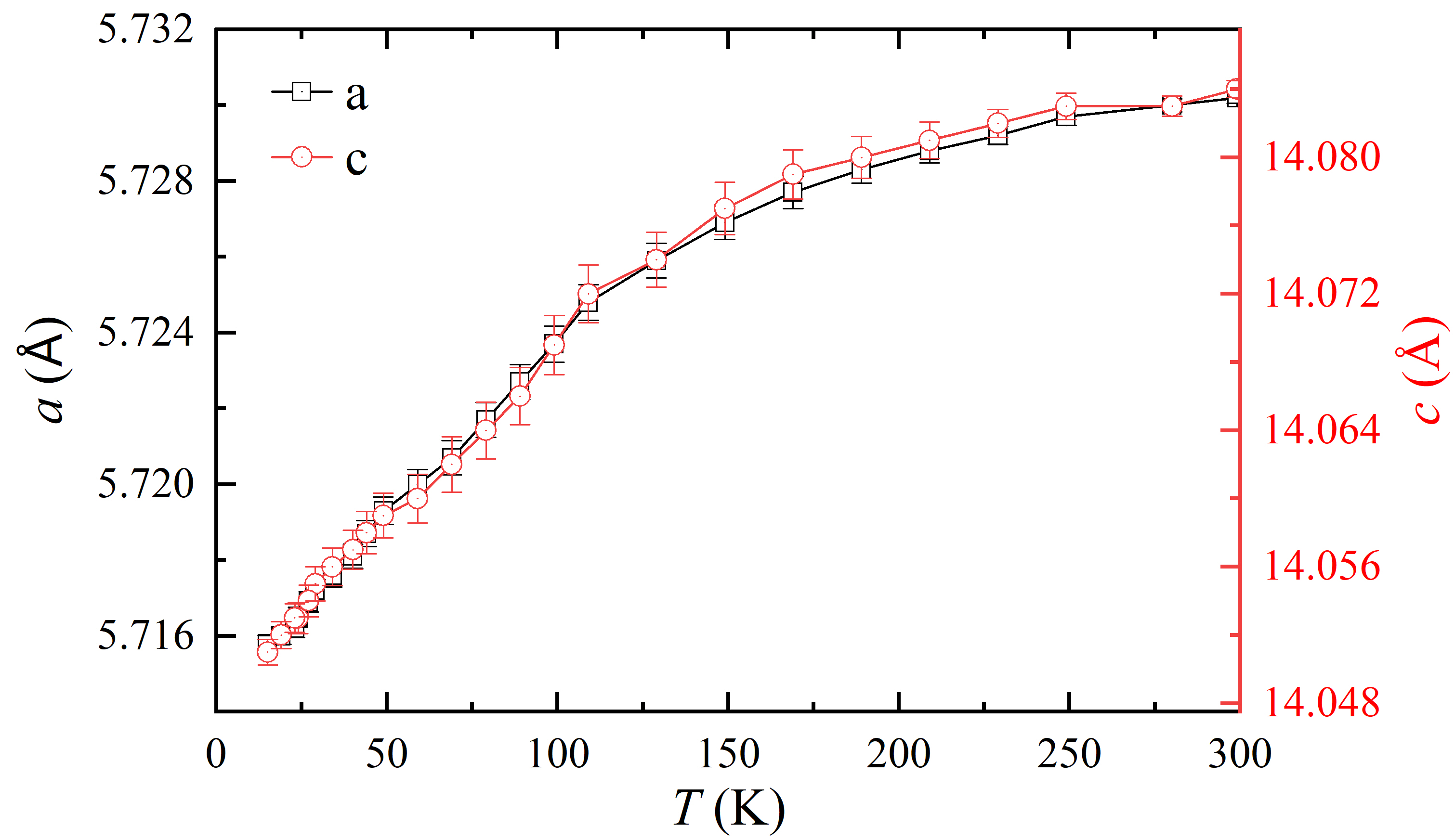}
		\caption{Temperature dependence of the lattice parameters \(a\) and \(c\) for \BSCO. Both parameters show a gradual decrease with cooling, indicating thermal contraction without any structural phase transition.}
		\label{Fig12}
	\end{figure}
	
	We performed temperature-dependent X-ray diffraction measurements on \BSCO to investigate possible structural changes associated with the magnetic transitions. The evolution of the lattice parameters, \(a\) and \(c\), was tracked from 300 K down to 15 K, as shown in Fig.~\ref{Fig12}. The extracted lattice parameters are in good agreement with those obtained from our neutron powder diffraction data.
	
	Our analysis reveals a smooth, monotonic decrease in both lattice constants with decreasing temperature, consistent with typical thermal contraction driven by anharmonic lattice vibrations.  Importantly, no anomalies or discontinuities are observed in the lattice parameters, indicating the absence of any structural phase transition in the investigated temperature range.
	
	The lack of structural distortions suggests that the observed spin-glass behavior in the magnetic measurements is not associated with a symmetry-lowering structural transition. Instead, the spin-glass state likely originates from intrinsic magnetic interactions, such as frustration arising from the face-sharing CoO$_6$ octahedral dimers or disorder effects due to Sb/Co site mixing. The absence of a structural anomaly further implies that the observed low-temperature magnetic dynamics are governed primarily by electronic and magnetic degrees of freedom rather than lattice distortions.
	
	\section{Discussion}
	\label{Dis}
	
	The comprehensive set of measurements—including DC and AC magnetization, heat capacity, magnetic memory and aging, electrical resistivity, temperature-dependent X-ray diffraction, and neutron powder diffraction—collectively establish that \BSCO\ does not exhibit long-range magnetic order down to \SI{1.5}{\kelvin}. Instead, it enters a frozen magnetic state consistent with a cluster spin-glass. Central to this behavior is the underlying crystal structure, which promotes disorder and magnetic frustration.
	
	Structurally, \BSCO\ crystallizes in a 6H-hexagonal perovskite structure featuring face-sharing $\mathrm{CoO_6}$ octahedral dimers. Each unit cell contains two crystallographically equivalent Co--Co dimers, with an intradimer Co--Co distance of approximately \SI{2.64}{\angstrom} and the shortest interdimer distance around \SI{5.73}{\angstrom}. The combination of these compact Co--Co dimers, significant intersite disorder between Co and Sb ions (\SI{\sim30}{\percent}), and notable oxygen non-stoichiometry (\(\delta \sim 0.54\)) introduces strong spatial variations in local magnetic exchange pathways. This structural disorder frustrates coherent long-range magnetic interactions and facilitates the formation of magnetically correlated clusters with short-range interactions.
	
	We expect that the Co--Co dimers exhibit predominant antiferromagnetic coupling within the dimers, due to direct exchange and orbital overlap effects inherent to face-sharing geometry. In contrast, the relatively larger interdimer distances and the observed positive Curie-Weiss temperature (\(\theta \sim \SI{21}{\kelvin}\)) suggest net ferromagnetic interactions between dimers within the \(ab\) plane. Similar magnetic motifs—antiferromagnetic intracluster coupling coexisting with ferromagnetic intercluster interactions—have been reported in other hexagonal perovskite systems, such as $\mathrm{Ba_4NbMn_3O_{12}}$~\cite{Streltsov2018}. This hierarchy of interactions—dimer-level antiferromagnetic and cluster-level ferromagnetic coupling—leads to a frustrated network of interacting clusters that collectively freeze into a glassy magnetic state.
	
	DC magnetization data reveal a clear bifurcation between zero-field-cooled and field-cooled curves at low temperatures, characteristic of spin-glass behavior. The high-temperature susceptibility follows a Curie-Weiss law, and the derived effective moment is consistent with a mixed-valence scenario for cobalt. Based on structural analysis and bond-valence considerations, we estimate that the Co ions adopt an average valence of +2.9, corresponding to a mixture of roughly \SI{10}{\percent} $\mathrm{Co^{2+}}$ and \SI{90}{\percent} $\mathrm{Co^{3+}}$. While our working model assumes high-spin $\mathrm{Co^{2+}}$ and intermediate-spin $\mathrm{Co^{3+}}$ states, we stress that these assignments remain tentative due to the sensitivity of cobalt spin states to local structural environments. In systems such as $\mathrm{LaCoO_3}$~\cite{Korotin1996, Maris2003, Haverkort2006}, even minor structural distortions are known to stabilize intermediate-spin states, and similar effects may be active in \BSCO.
	
	AC susceptibility data exhibit a frequency-dependent shift in the freezing temperature, and the dynamics are best described by the Vogel-Fulcher model. This points to a collective spin-freezing process involving weakly interacting spin clusters, rather than independent paramagnetic spins. Magnetization isotherms further support this interpretation: their nonlinearity at low temperatures and the extracted effective moment from Langevin fitting are consistent with the presence of correlated spin clusters.
	
	The non-equilibrium nature of the magnetic state is corroborated by thermoremanent magnetization and magnetic memory experiments, which display pronounced aging effects and history dependence—key signatures of spin-glass dynamics. Heat capacity data also align with this picture, showing no sharp anomalies indicative of magnetic ordering but instead a broad hump at low temperatures, likely arising from short-range spin correlations and low-energy excitations.
	
	Temperature-dependent XRD and neutron powder diffraction confirm that no structural phase transition occurs across the magnetic transition. The continuous lattice contraction and unchanged space group symmetry support the conclusion that the spin freezing is driven not by a structural instability but by intrinsic frustration and disorder. Neutron diffraction further reveals no long-range magnetic order or significant diffuse scattering, implying that the correlation length of magnetic clusters is short and the signal is buried in the background.
	
	Finally, electrical resistivity measurements show semiconducting behavior with Arrhenius-type activation and no evidence for variable-range hopping, implying thermally activated transport and further underscoring the role of disorder in both electronic and magnetic properties.
	
	In summary, the structural framework of \BSCO—characterized by face-sharing $\mathrm{CoO_6}$ dimers, substantial cation disorder, and oxygen vacancies—plays a pivotal role in promoting frustrated interactions and cluster formation. These features culminate in a cluster spin-glass ground state with slow dynamics and emergent magnetic memory effects. Future studies using local probes such as muon spin relaxation (\(\mu\)SR) or inelastic neutron scattering will be instrumental in revealing the microscopic origin of these correlated clusters and their low-energy excitations.

	\section{Conclusion} 
	\label{Con}
	
	Our comprehensive study of \BSCO reveals a low-temperature state devoid of long-range magnetic order. DC and AC susceptibility measurements point to a spin-freezing transition with Vogel–Fulcher dynamics. At the same time, Langevin fits of the isothermal magnetization indicate the formation of finite-sized, interacting spin clusters. Complementary thermoremanent magnetization and memory effect experiments reinforce a spin-glass-like state. Heat capacity and neutron diffraction data show no sharp anomalies or magnetic Bragg peaks down to 1.5~K. Rietveld refinements underscore significant Co/Sb intersite disorder and oxygen vacancies, likely introducing exchange randomness and magnetic frustration. Additionally, our findings further support the semiconducting behavior observed via an Arrhenius-type resistivity and the absence of structural transitions from temperature-dependent XRD.
	
	These results establish \BSCO as a promising cluster spin-glass candidate in which intrinsic disorder and frustration suppress conventional magnetic order. The exotic spin state of Co in face-sharing hexagonal perovskites thus emerges as an intriguing platform for future studies. Probing this state with advanced local and spectroscopic techniques may unveil novel quantum phenomena and inspire innovative approaches to tailor spin dynamics in complex oxide systems. Future studies on related hexagonal perovskites with controlled Co/Sb ordering or alternative dimer-forming ions (e.g., Rh, Ir, or Ru) may offer further insight into how structural motifs and disorder levels govern the emergence of glassy versus quantum-disordered ground states.
	
	\section*{Acknowledgments}
	
	We gratefully acknowledge Eva Brücher for her valuable assistance with magnetization, heat capacity, and resistivity measurements using the Quantum Design MPMS-XL and PPMS systems at the Max Planck Institute for Solid State Research. We also thank R. Rae and C. Kirk for their support with the X-ray diffraction measurements at the University of Edinburgh, UK. We are further indebted to C. Stock (School of Physics and Astronomy, University of Edinburgh) for insightful and stimulating discussions.
	
	\medskip
	\noindent \textbf{Note:} AA and GK contributed equally to this work.

	\bibliography{Ref}
	
	\appendix
	
	\renewcommand{\thefigure}{A\arabic{figure}}
	\renewcommand{\thetable}{A\arabic{table}}
	\setcounter{figure}{0} % Reset figure counter
	\setcounter{table}{0}  % Reset table counter
	
	\section{Temperature-Dependent XRD and NPD}
	
	\begin{figure}
		\centering
		\includegraphics[width=1.0\linewidth]{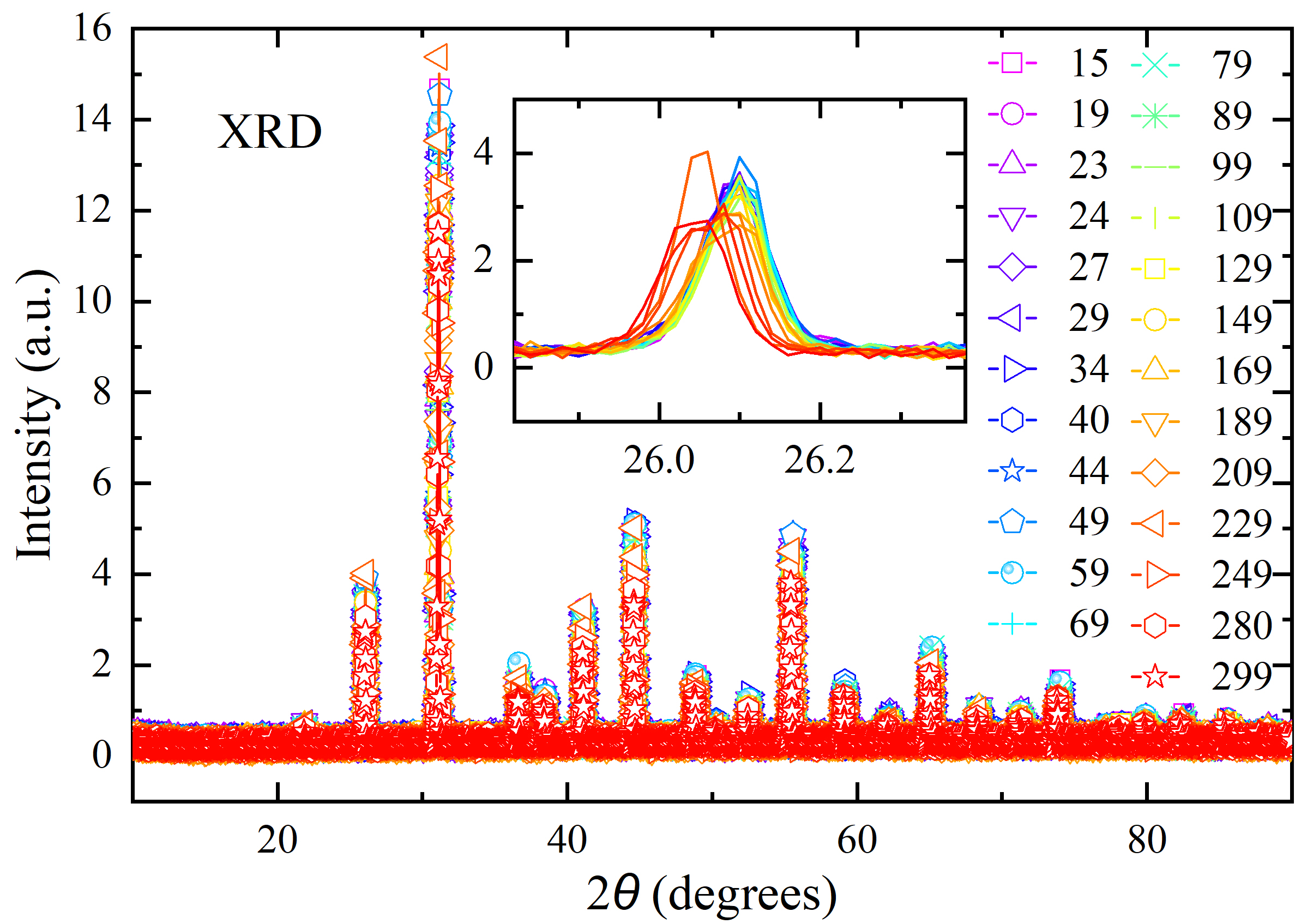}
		\caption{Temperature-dependent X-ray diffraction patterns of \BSCO measured from 15~K to 300~K. The inset shows an enlarged view of a selected peak, highlighting the shift in peak position with temperature.}
		\label{A1}
	\end{figure}
	
	\begin{figure}
		\centering
		\includegraphics[width=1.0\linewidth]{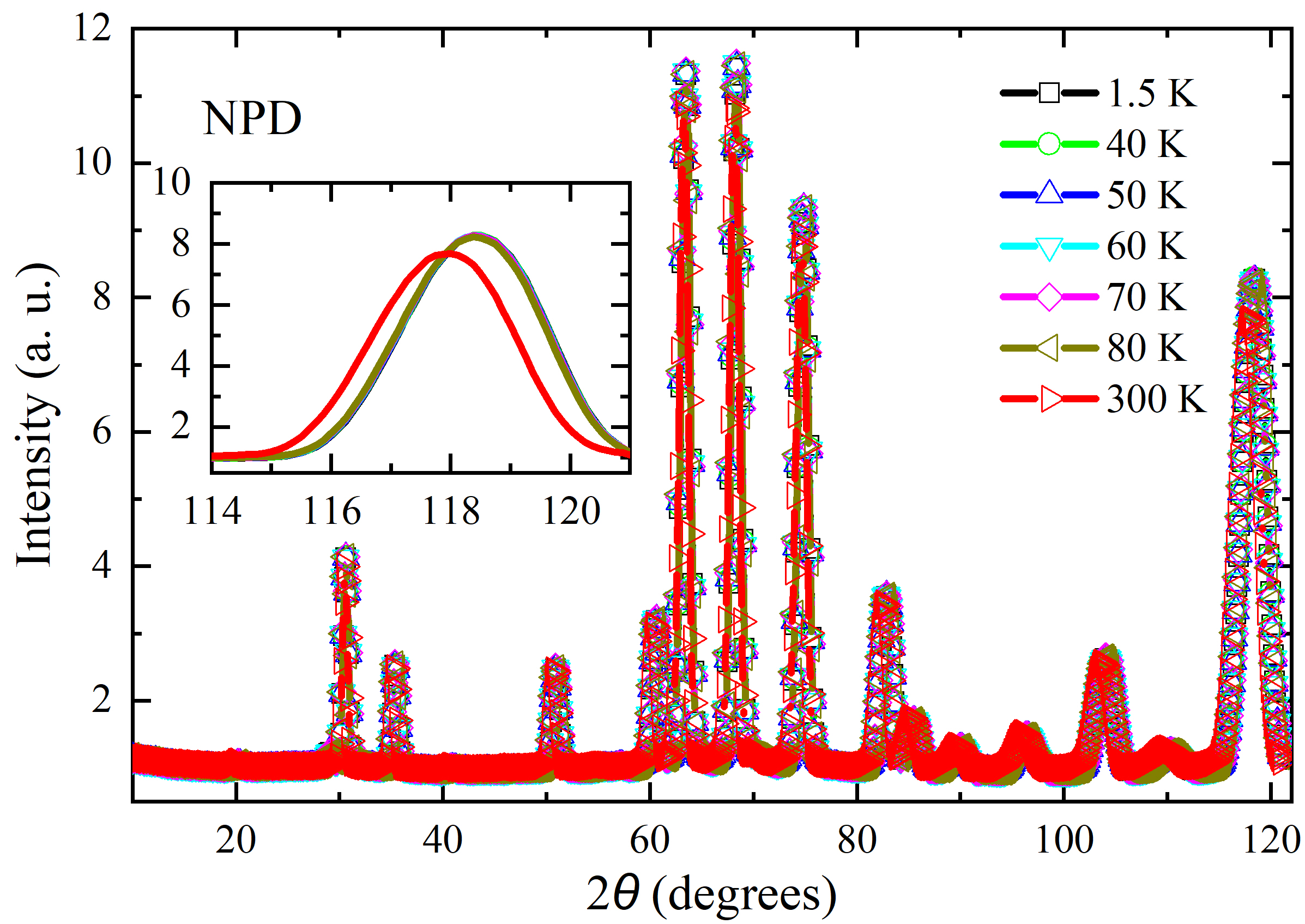}
		\caption{Temperature-dependent neutron powder diffraction patterns of \BSCO measured from 1.5~K to 300~K. No additional reflections appear at low temperatures, confirming the absence of magnetic transitions. The inset presents an enlarged view of a selected peak to illustrate the temperature-dependent shift in peak position.}
		\label{A2}
	\end{figure}
	
	To investigate the structural stability of \BSCO as a function of temperature, we performed temperature-dependent X-ray diffraction and neutron powder diffraction measurements from 1.5~K to 300~K. The XRD and NPD patterns exhibit no additional diffraction peaks at low temperatures, indicating the absence of structural and magnetic phase transitions. Instead, a systematic shift in peak positions with decreasing temperature suggests thermal contraction of the lattice. The insets of Figures~\ref{A1} and \ref{A2} highlight an enlarged view of a representative diffraction peak, illustrating this shift in more detail.
	
	\section{Bond valence sum}
	
	\begin{table}
		\centering
		\caption{Bond valence sum calculations for \BSCO\ at room temperature, extracted from Rietveld refinement of neutron diffraction data collected at room temperature. The BVS values were determined using the \textsc{EXPO Version 2.3.5} software to estimate the effective charge on each atom based on the assumed valence states. CN denotes the coordination number.}
		\label{BVS}
		\setlength\extrarowheight{4pt}
		\setlength{\tabcolsep}{10pt}
		\begin{tabular}{cccc}
			\toprule
			Atom & CN & Valence Assumed &  BVS  \\ \midrule
			Ba1  & 12 &       +2        & 2.254 \\
			Ba2  & 12 &       +2        & 2.175 \\
			Co1  & 6  &       +3        & 2.313 \\
			Sb2  & 6  &       +5        & 4.822 \\
			Sb1  & 6  &       +5        & 4.496 \\
			Co2  & 6  &       +3        & 2.074 \\
			 O1  & 8  &       -2        & 1.878 \\
			 O2  & 8  &       -2        & 1.874 \\ \bottomrule
		\end{tabular}
	\end{table}
	
	The bond valence sum analysis, performed using \textsc{EXPO Version 2.3.5}~\cite{EXPO} on neutron diffraction data, provides insights into the local bonding environment and charge distribution within \BSCO. The BVS values for Ba atoms are close to the expected +2 valence, while those for Sb atoms are slightly below the ideal +5, suggesting minor structural distortions or covalency effects. The Co sites exhibit deviations from the assumed +3 valence, with Co1 showing an underbonded state (2.313) and Co2 being closer to +2 (2.074), supporting the presence of mixed-valence Co$^{2+}$/Co$^{3+}$ states inferred from magnetic measurements. The oxygen atoms display BVS values slightly below the ideal -2, consistent with oxygen non-stoichiometry observed in the refinement. These results highlight the interplay between cation disorder, oxygen vacancies, and the complex charge distribution, which likely contribute to the unconventional magnetic behavior of the compound.

\end{document}